\documentclass{article}
\usepackage{amsmath}

\usepackage{amsfonts}
\usepackage{amsthm}
\usepackage{graphicx}

\begin{document}

\title{Informal Control Code Logic}

\author{Jan A. Bergstra\thanks{This work has been carried out in the context of NWO Jacquard project Symbiosis.}
\\ 
\newline\\
University of Amsterdam, Faculty of Science, Informatics Institute, 
\\Section Theory of Computer Science, \\
Science Park 904, 1098 XH, The Netherlands\thanks{P.O.~Box~94214, 1090~GE~Amsterdam,  The Netherlands E-mail: \texttt{j.a.bergstra@uva.nl}}
}

\date{September 12, 2010}

\maketitle

\begin{abstract}
General definitions as well as rules of reasoning regarding control code
production, distribution, deployment, and usage are described. The role 
of testing, trust, confidence and risk analysis is considered.
A rationale for control code testing is sought and found, for the case of
safety critical embedded control code.

\end{abstract}

\section{Introduction}
\label{sect-intro}
Although the combination of hardware and software into functional units pervades 
computing technology the question why
this organizing principle of information technology is so powerful and why it has 
emerged during the evolution of digital equipment is intriguing, if only 
from a theoretical viewpoint. Following the terminology of \cite{BMCCL} software is understood as 
control code and data code.
In this paper a machine is called a box. A control code logic (CCL) provides formal reasoning methods that apply to
control code
and code controlled boxes. According to \cite{BMCCL} conditional equations constitute a workable formalism for a 
CCL.\footnote{A brief summary of \cite{BMCCL}: (i) control codes are either produced via instruction sequences by 
means of compilation or differently, for instance machine learning (such control codes are termed dark programs 
in \cite{Janlert2008}), (ii) no
substantial definition of `executable code' can be found, rather executability needs to be introduced as a primitive predicate
together with a machine model, (iii) some basic facts about compilers and assemblers can be phrased in terms of 
conditional equations (in particular several forms of fixed points), (iii) a formalization of some arguments from \cite{Appel1994},
\cite{Halpern1965}
and \cite{EarlySturgis1970}
 (iv)  fairly detailed definitions of installation and portability
are given in terms of CCL only.}
Informal control code logic (ICCL)\footnote{In  \cite{HorlickJones2005}  another 
occurrence of a specialized informal logic is presented, and \cite{Hansson2004} discusses (fallacious) 
reasoning rules in the same field.} provides informal reasoning patterns about the use of control codes.

Many ways to approach this kind of question can be imagined. The history of technology can be followed in detail,
conjectural histories may be developed (quite common in the theory of money, see \cite{Bergstra2010a}), or
economic models may be construed from which the advantages and disadvantages of technological strategies
can be analyzed.

None of these approaches are available to us at this stage because of the lack of a general conceptual model of the 
situation. The primary objective of this paper is to develop that conceptual model. It seems to be the case that given such 
an objective 
one needs to proceed from scratch. I will proceed in a holistic way by discussing a number of aspects:
\begin{enumerate}
\item I will make use  of imaginative definitions as presented in \cite{Bergstra2010a}, and in particular the 
notions of an IDBR (informal description by role), an LSCD (logical  solitary concept definition) and an SCFD
(stratified concept family definition). An IDBR for a code controlled device (also  called a box) is provided
in the line of the machine function based definitions of \cite{BMCCL}. 
\item I will make use of synthetic execution architectures  as defined in \cite{BPEA} and of a
generalized form of the analytic execution of  \cite{BPEA}.
\item Boxes are instances of box types. Box types develop along a path of technological evolution. 
Within a type a sequence
of versions (subtypes) may be distinguished. Control codes can be exchanged between boxes and
information carriers. A classification of exchange mechanisms is proposed.
\item A variety of control related actions are brought into the picture: 
\begin{itemize}
\item shipping control code, (thrashing control code),
\item loading control code, (unloading control code), 
\item decision to use a box, (decision not to use a box), 
\item usage (of a box) = carrying out the decision to use a box
\item idling (of a box) = carrying out the decision not to use a box
\item utility (when making successful and intended use of a box), 
\item disutility (when making use of a box without intended success), 
\item expected utility, (ex ante assessment),
\item (expected disutility), 
\item satisfycing usage, (dissatisfycing usage), 
\item $[$user side/producer side$]$ control code testing, 
\item producer side control code validation, 
\item producer side control code verification.
\item user side control code deployment risk analysis
\item producer side control code release risk analyisis
\end{itemize}
\item The primary category to which all other actions of agents, including users and producers, is that of decision making. 
Testing, risk analysis, reputation assessment, confidence awareness, and trust maintenance all contribute to decision making.
It is assumed that under normal conditions the use of a control code does not take place without a preceding decision to to so.
\end{enumerate}

Each of these aspects plays a role in the construction of an ICCL level of abstraction for describing computer technology. Perhaps
too many aspects have been collected below and a more concise discussion can be developed. In preparation of the discussion I mention two 
philosophical perspectives that somehow threaten and potentially overshadow my project in each of it stages.
\subsubsection{Looking for meanings is futile}
I am tempted to ask questions like: ``what is a control code,', ``what is a code controlled device'', ``what is testing a control code'', 
``what is a computer'' and so on. Another style of phrasing would be: `what is the meaning of ``control code'' ', 
`what is the meaning of  ``code controlled device'' ', `what is the meaning of  ``testing a control code'' ', and
`what is the meaning of ``computer'' '. Now in \cite{Alston1963} Alston expands in long detail on the impossibility of finding such meanings.
He considers the search for meanings futile because there is no general class of ``meanings'' within which one is searching. As a consequence
I will choose this strategy: instead ask: `what is a definition  of ``control code'' ', 
`what is a definition of  ``code controlled device'' ', `what is a definition of  ``testing a control code'' ', and
`what is a definition of ``computer'' '. And in order to escape from Alston's critique a meaning needs to be assigned to this use of the term
``definition''. That is done by making use of a classification of definitions  taken from \cite{Bergstra2010a}. Nevertheless it is easy to ask for 
meanings of seemingly straightforward terms and phrases and to underestimate the force of Alston's critique. Alston formulates a difficulty that each
quest for meaning needs to confront.

When writing about such meanings work of previous authors needs to be used. Years ago Jervis \cite{Jervis1980}  formulated the following 
criticism (then concerning some books he was reviewing): ``Although there are lots of footnotes and references, in most cases these are the obligatory 
nods that substitute for refining and building on what others have done.''  Avoiding this problem is a very difficult objective indeed and I am very reluctant
to claim successfully having done so.

\subsubsection{NOMA}
As I will discuss in more detail below testing can produce knowledge about control code and in particular about the behavior of a 
code controlled box upon its execution. This knowledge may consists of  reproducible data about physical phenomena. Quite at the opposite end of
the spectrum one may envisage a control code which has been obtained via compilation of an instruction sequence which is supposed to be 
executed by a box for which a formal model of execution (in terms of the underlying instruction sequences) is given. 
Then predictions about behavior can be made with mathematical (logical) precision.

Unclear is to what extent both forms of knowledge can coincide or overlap. It is almost
inconceivable that pure formal reasoning can produce a valid
prediction about the behavior of a physical device. Somewhere the outcome of formal reasoning must be combined with experimental data to 
arrive at a prediction of observable quantities. Thus experimental data (obtained via control code testing) and logically derived information seem to 
exist in different worlds. 

Whether or not these worlds actually differ is left open below. I will in some cases assume that there may be an alternative to testing for 
obtaining certain information. Whoever thinks that testing is insufficiently reliable for control code validation must believe that such 
alternative methods for obtaining information can be found, at least under certain conditions. Here lies a philosophical issue that will not be 
further analyzed below, in spite of the fact that its resolution seems to be rather essential for our objectives.

If one concedes that both methods of knowledge acquisition lead to disjoint sets of possible results one may wonder whether or 
not this form of seemingly related but  in fact non-overlapping methods of investigation constitutes new circumstances in terms
of the philosophy of science. Such is not the case, however. Following \cite{Moritz2009} an example of  non-overlapping theories is found when
contemplating the results of science in contrast with the results of theological reflection. Moritz, in  \cite{Moritz2009} discusses NOMA: 
non-overlapping magisteria of authority (NOMA), a concept put forward by the ethologist S. J. Gould.  Moritz concludes, however, that NOMA 
inadequately describes the relation between science and theology because according to him overlaps between these areas cannot be excluded. 

An asymmetry between positive and negative information may be at hand, however. 
In the case of control codes, confidence generating information obtained from tests may be disjoint
from confidence generating information obtained by means of logical analysis while logical analysis might lead to a confidence loss 
(negative information) which subsequently
suggests useful tests (the outcome of which may confirm the same negative information). 
This state of affairs is comparable to the fairly detailed NOMA related viewpoint put forward by Stenmark in \cite{Stenmark2005}.
Making a connection between ICCL and this perhaps marginal part of philosophy of science may seem quite farfetched. But the presence
of some form of NOMA, though regarding quite different areas of authority, is not entirely farfetched. 
The huge gap between those who insist on software testing and those who insist on formal software verification, both groups of 
individuals often not trusting one-another's methods, is a stable and significant phenomenon which calls for a further explanation 
beyond the suggestion that the methodology of formal verification is still largely unfamiliar to most who have been used to testing.

\subsection{The ICCL level of abstraction}
If computers and computer systems are considered from an ICCL perspective that means first and for all that no
hypothesis is made and no understanding is assumed in any form concerning the mechanical aspects of control code
execution.\footnote{In particular the common explanation that this is a matter of computer program execution is 
rejected as being uninformative.}

To what extent an ICCL perspective on computing can be maintained is unclear. Further development of ICCL may extend
the frontiers of ICCL and reveal its intrinsic boundaries so to speak. 
It is considered to be consistent with ICCL to make assumptions about the
bookkeeping and management of bit sequences within a system. Below that will hardly be done, but
explaining the operating architecture of a system in terms of a number of self-explanatory file management primitives (commands)
is needed
for an investigation of (i)  the simultaneous presence of control codes for different applications
within the same system, and (ii) of the distinction between control code ad data code as well as (iii) the way data code is 
preserved between sessions of execution of the same control code.

Instruction sequences can be used to specify threads for issuing commands to an operating architecture. In \cite{BMCCL} a
 proposal for an
operating architecture is formulated. Threads are specified that embody definitions of control code installation and 
control code portability. Similar specifications can be found in \cite{BerWal03}.

A major incentive for this work has been to develop some theory about the business cases for software process outsourcing.
That led to the objective to disentangle that issue entirely from the endless complexities of control code execution mechanics and
control code structure.
This application perspective is not pursued  below and from the standpoint of software process sourcing (or software asset sourcing
as it has been named alternatively)  only preparations are made. It seems clear, however, that significant progress concerning
operating architecture description is needed if significant conclusions regarding in- and out-sourcing decisions are to be
drawn from consideration of the matter at the abstraction level of ICCL.
\subsection{Setting the scene}
The term satisficing usage is coined following the terminology of H.A. Simon for decisions in the context of 
decision making.
Having available this terminology we provide a number of preparatory remarks.
\begin{enumerate}
\item We may assume that a box B loaded with control code C creates for its user U (when used at its initiative) 
either a utility (UT) or a disutility (dUT). 
The only way that UT, provided it is non-trivial (that is nonzero),
can come about given that C and B are both at hand for U is by U's executing C on B.
The same holds for dUT if it is nonzero. It is the effect of a failed execution which is of a lesser utility than 
not to execute C on B at all.\footnote{Not executing C on B may deliver some disutilities as well but it can 
be done without having C or B or both C and B at hand. But an unsuccessful execution of C on B may create a 
disutility which is far
more problematic than not executing C at all if during the execution of C before a failure took place a new 
state has been formed from which recovery is only possible by making use of the tail of an execution of C on B.}
Here it is assumed that no other comparable boxes and codes are ready for use by U as an 
alternative.\footnote{Thus for instance mere code inspection of C by U cannot bring forward either the 
mentioned utility or the mentioned disutility because it fails to involve an execution of the code.}

\item Usage (of a box, or of a control code for a box) involves the concept of execution of a box. 
This implies that when writing about it a box on which the usage is performed needs to be so `real' that it can 
be executed. But of course it isn't. So here is a key difficulty for writing about usage: 
one intends to write about effects only achieved by execution while writing about abstractions that 
cannot possibly execute.

\item Usage is a concept with several close relatives: for instance
experiment, demonstration and test (including validation). Except interpretation and simulation
each of these variations on the theme of usage involves an execution of the box. 
But the test provides none of the utilities that usage provides and instead it provides quite different
utilities (if at all). A test cannot deliver a disutility (though a false positive comes close). 
The following variations of usage can be distinguished.
\begin{description}
\item{\bf{Satisficing usage.}} Practical activity performed by one or more users (sometimes hierarchically stratified in
supporting users and end users al of whom are considered to be human beings) centered on one or more runs of the box, 
performed, to U's satisfaction, to meet U's objectives, as embodied
in the utilities of usage.
\item{\bf{Usage.}} Practical activity essentially involving one or more runs of the box, performed to 
meet U's objectives, as embodied
in the utilities of usage (and with failure captured by the disutilities of usage).
\item{\bf{Interpretation.}} An interpretation is a simulation which is performed by way of usage. 
So it may deliver utilities close to UT or disutilities close to dUT.
\item{\bf{Demonstration.}} Demonstration also involves box runs and its outcome is knowledge for the user or its 
representatives. 
\item{\bf{Test.}} Practical activity involving one or more runs of the box, that will not deliver any of the utilities or disutilities
of usage, but instead it will provide knowledge about the box. This knowledge is in principle to be used for decisions about
usage or about shipping of boxes and/or control code.
\item {\bf{Experiment.}} An experiment involves runs of the box. Its outcome is a hypothesis about box behavior. 
Such hypothesis may be further examined by way of test. Inductive inference leads from experiment to hypothesis.
\item{\bf{Simulation.}} A simulation does not involve a run of the box. It involves  run's of a modified control code on a 
modified box (at least one of the modification is non-trivial or else the simulation is just an experiment.) No 
usage utilities (disutilities) will result.  The objective of a 
simulation is hypothesis formation (concerning the target box) to be further reinforced by means of experiments and 
thereafter examined by way  test.
\end{description}
\item The viewpoint that tests require a hypothesis for confirmation or for disconfirmation while experiments may 
generate a hypothesis
is incongruent with the experimental sciences where the above `test' is rather termed an experiment.
\end{enumerate}
 There is a plausible 
order of events as follows: once a control code has been produced its simulation on a machine different from a 
target machine may generate hypotheses about its functionality. Subsequently an experiment on the target boxes may 
reinforce these hypotheses,
subsequently the hypotheses are subject to tests. If passed demonstrations are due, followed by usage and upon a positive
evaluation of the usage a phase of satisfycing usage may follow.

\subsection{Control code exchange}\label{CCEM}
The crux of code controlled boxes is that control codes contained in a box can be exchanged. 
``Box with exchangeable control code'' is 
another phrase for ``code controlled box''.\footnote{Alternatively the feature of a box that it admits
exchangeable control code is termed the exchangeable control code feature, or simply the control code feature.}
For box users the advantage of exchangeable control  code must
be connected to their performing the act of control code exchange at least in some cases. More precisely a user
obtains an advantage from the mere possibility of control code exchange by being certain that it can be done even if
that is so rare that most users never make use of a control code exchange. The minimal role of the control code feature is
to serve as an insurance policy for a box owner that his box can be made useful for other purposes when needed.

This implies that in order to assess the feature of exchangeable control code for boxes some clarity about the 
exchange mechanisms must be provided. It is taken for granted that exchanging control code can be viewed as
a combination of dropping a loaded control code which is considered unproblematic and loading another control code.
As it turns out several scenarios exist for loading new control codes into a box. 
Here are some options:\footnote{Besides ranging from technologically simple to quite intricate, this listing corresponds
with a historical development in technology as well. This paper not being about history of computing it may be safer to
attribute the listing some virtue concerning the chronology of a conjectural history, 
perhaps one may even speak of evidence based conjectural history.}
\begin{description}
\item{\bf{Plugged external memory device (PEMD).}} This refers to a CD (in former day a floppy disk), or a memory stick, 
which is connected directly to the box without any network in between. Controls on the device allow downloading data 
(bit sequences often called files) from the PEMD and  uploading  data from the box to the PEMD.
\item{\bf{Wired EMD.}} An external disk drive is an example: the EMD has some functionality 
usually requiring independent power supply; once a cable connection has been established between EMD and box 
exchanges can be controlled from the box's user interface.
\item{\bf{Wireless EMD.}} Connection with an EMD may be established via bluetooth or a comparable technology which 
does not presuppose full internet connectivity.
\item{\bf{Wired internet connection.}} A statically positioned box may have IP 
capability and be internet connected via a wire.
\item{\bf{Wireless internet connection.}} This comes in several flavors. 
\begin{description}
\item{\emph{Very local area network.}} At the smallest scale a bluetooth connection may connect
the box to an internet host from a small distance (several meters). 
\item{\emph{Local area network.}} At a small scale (say up to 20 meters) a network may
provide wireless link to a local host, without options for roaming. 
\item{\emph{Wide area network.}} At a larger scale a system of base stations may provide
connectivity from the box throughout a wider area, say a campus comprising some square kilometers.
\item{\emph{Global network.}} World wide connectivity may be provided if the box contains a mobile phone.
\end{description}
\end{description}
Loading a new or improved CC requires conscious user activity if a PEMD is used, and it may also require conscious 
action activity if wired or wireless EMD is used. All methods from Wired internet connection onwards are said to be 
internet based.

\subsubsection{Network based exchange implies exchangeable control code.}
Remarkably the issue of determination the rationale (or different rationales) of exchangeable control code 
as a feature of box technology
 is far from independent from these different options for control code exchange. If PEMD is used (for instance a CD disk,
or historically a floppy disk)
it is certainly conceivable that a box performs tasks that may as well be performed by boxes of different design 
that are not code controlled. But if a box has global wireless
internet connectivity via the mobile phone network, it is most plausible to assume that its processing capabilities are 
so advanced that it must be code controlled and if it is not, that option has been blocked in hindsight after a 
satisfactory  code controlled  prototype had been constructed during the design and development of the box. 

Assuming that machines are more likely to be code controlled with growing functionality\footnote{This assumption
itself reflects limitations. Human brains are complex but seem not to be code controlled.} it appears that the question 
why code controllability is so important is mainly meaningful if simple methods of CC exchange are considered such 
as PEMD, wired EMD and wireless EMD. Considering current and past technology one may agree that  wireless 
global area network connectivity
is present only for boxes for which the control code feature is plausible which for that reason is available 
by default, and for which its absence rather than its presence is expected to be specifically engineered, for instance 
for security reasons.

\subsubsection{Hidden versus visible control code exchange.}
The observation that the control code feature's plausibility depends on assumptions about code exchange mechanisms
 leads to the proposition that exchangeable control code comes in different flavors: 
\begin{description}
\item{\emph{visible control code feature}}: control code exchange is only performed as an immediate consequence 
of conscious user activities. Such activities are classified as configuration management by hand, or equivalently
as interactive configuration management.
\item{\emph{hidden control code feature}}: changes made only by autonomous system actions either initiated from
an external network component or from the box at hand but in both cases hidden from a user. This is the
hidden control code feature, it involves automatic configuration management. Of course if many
control codes are simultaneously loaded a mixture of automated and interactive configuration management 
can be in charge for control code exchange on the same box.
\item{\emph{illegal control code feature}}: if control code exchange can be achieved and in addition it can only be achieved 
by a conscious but illegal or unintended (from the perspective of the box provider) interactive configuration 
management by the user the box offers an illegal control code feature. If control code
has been exchanged under these conditions the box is said to have been hacked. Of course mixtures with the 
visible CC feature and the hidden CC feature can be imagined.
\item{\emph{illegal hidden control code feature}}: this applies if automated activities can lead to control code exchanges that
are illegal or unwanted from the perspective of the box provider (and in most cases also from the perspective of the box
user). Viruses and worms, Trojan horses and more general malicious control code (malcode in brief)
usually enter a box in this 
fashion.\footnote{In some cases a control code exchange that imports a malicious code will require user activity while
the user is unaware that this is a consequence of his actions. Then we may speak of a user mediated hidden 
control code feature. It is implicit that the user is unaware of the mediating role of his actions which most likely are
performed with another purpose, with opening an email attachment as a paradigmatic example.}
\end{description}

\subsubsection{Conceptual questions about the control code feature.}
The question to explain the large number of code controlled devices seems to split into different parts:
\begin{enumerate}
\item Why is it the case that beyond a certain level of complexity and functionality all boxes are code controlled?
\item What explains the presence or absence of the code control feature for box types with functionalities that do not 
by their own complexity indicate the code control feature as the most plausible default? (For instance: why
do digital photo cameras not offer the control code feature?)
\item Concerning the concept  of computer science and its methodology several questions  emerge: 
\begin{itemize}
	\item is a computer a box with code control feature by default or is that merely and option?
	\item If a computer is by default code controlled is it supposed to be universal in some sense?
	\item If computers are considered not to be code controlled (and for that reason not to be universal either), 
	what is the characteristic subject of computer science which provides its conceptual identity?
	\item If the latter questions have unsatisfactory answers due to the lacking expressive power of CCL: 
	can computer science more convincingly be defined as the science of tasks 
	that can be performed by code controlled boxes with codes being produced by compiling instruction sequences.
\end{itemize}
\item How to express in terms of code controlled boxes that a control code is malicious? How to express (if that can be done)
that a malicious control code is a virus, a worm, a Trojan horse and whatever other kind of malcode has been
distinguished?
\end{enumerate}

\section{Control code related decision making}

I will assume that all actions involving control code are performed by conscious agents and that each action is preceded by a decision tot that extent.
The decision to use a specific control code can be an involuntary one in the sense of \cite{Haavelmo1950}, for instance if 
malicious code is being activated.
That case will not be dealt with below. Decision making concerning control code can vary from a slow and strategic process involving many agents 
and moving through a pattern of formalized steps as if a sequential program were executed (following H.A. Simon's description of decision making, 
see \cite{Simon1959,CyertEtAl1956}, and referred to as formality of a decision in for instance \cite{Avlonis1985}) 
to a real time action disallowing much opportunity for 
systematic reflection. A typical study on real time decision making is \cite{MahalelZaidelKlein1985} analyzing the decision taking behavior of traffic 
participants in front of traffic lights. This may be compared to internet users who may decide about control code activation each time 
they make up their mind about opening an email attachment. In this paper I will assume a highly informed decision making process, thus avoiding
the complications of empirical investigation of actual decision making processes hampered by
real time pressure and a correlated lack of information. This assumption is by no means realistic but it creates more idealized circumstances that 
may simplify the theoretical work. 

\subsection{Making the decision to use a specific control code}
Before U puts a control code into use by having it executed on box B for some specific purpose U will make a decision to 
proceed in that way. Devoid of cognitive insight in the structure of the control code he will need another basis for making the decision.

The user U of a control code primarily depends on his trust in the control code producer. This trust may be based on any combination of the following
matters: 
\begin{itemize}
\item Past experience with other control code products manufactured by the same producer.
\item Knowledge about the production process that has been applied. This knowledge can be based upon reputation analysis or upon 
detailed production process inspections performed by U or on behalf of U.
\item Reputation of the control code as a product obtained by analyzing the satisfaction of previous users.
\end{itemize}

Besides acquiring confidence that the control code is adequate based on trust in its producer users can perform tests
and users can obtain information about other instances of the same the control codes from other users they happen to know and in whom 
they have sufficient trust. 

Besides reputation based trust of its producer and trust based upon production process inspection testing is the only method 
available to a user to increase
his confidence that the control codes will lead to satisfycing usage. This dominant  role (and interaction) of reputation, trust, 
test and confidence is specific for a setting where ICCL is leading.\footnote{It seems plausible to place open source near the opposite end 
of this spectrum, where the extreme opposite end is occupied by control code production and release methods where 
besides a control code also a specification and a correctness proof which can be checked by the user is handed over from 
producer to user. If the check fails or if the specification is incompatible with user requirements the sale is off.}

So it can be concluded that when it comes to decision making 
ICCL assigns a central role to control code testing, and because of this central role  testing 
is investigated next in more detail.
The general question that needs to be answered is this: what knowledge about C can be generated by running B(C) with a 
number of different inputs. Control code testing takes place at least at two sides: user and manufacturer. Intermediate
agents like software salesmen, system administrators, IT consultants may perform tests as well, but these agents 
(or rather agent roles) will be ignored in the sequel of this paper.

\subsection{Decision making in  general}
Scanning the literature on decision making I get the impression that the foundational question ``what is a decision'' has not attracted much 
attention.\footnote{In contrast the literature on risk analysis pays much attention to ``what is risk'', and 
the literature on finance has many contributions
on ``what is money''. But in computer science one observes a similar phenomenon: questions like ``what is a computer program'', 
``what is an operating system'', ``what is computer program testing'', ``what is malicious code'', and ``what is a computer virus'',
do not attract significant attention.}
Yet for the current objectives it is a rather intriguing question to settle that matter in more detail. Below I will put forward
an elementary theory of decisions. It is not
supposed to have any empirical content. I assume that an agent or a collection of agents proceeds by performing actions of various kinds. These 
actions include observations of atomic propositions and all actions may have side-effects. A classification of actions has been made into 
operational actions henceforth called primary actions which directly constitute to the major performance of these agents and collateral actions 
henceforth called secondary actions which involve planning, communication, what is usually called decision making and so on.

When considering a trace of a system its secondary actions can be grouped together in threads of actions that are semantically connected. The secondary
actions take place as if a multi-threaded system proceeds through one of its strategic interleavings. Each of its threads may be understood as the execution
of some sequential process (or program in the terminology of \cite{Simon1959}). Focusing on one of the threads involved an external observer 
may or may not be justified in making the claim that this thread constitutes (represents or documents) the process of making some decision. It is
a decision if that claim is valid. Several grounds for validity of such claims can exist. Quite common is the situation that many of its actions are labeled with
comments about their being constituents of decision making process. For a ``formal theory'' of decision (formal decision process view) 
that criterion will suffice: a decision is a tread of 
actions which conforms one of the possible alternatives of a pre-existing grammar for decision making processes. This grammar will include label 
assignments of various actions that explain which particular stage of the decision process these actions incorporate.

Another view may be termed the ``decisive progression'' view. Here a progression is a sequence of actions (usually taken from a larger sequence), see \cite{BergstraPonse2009}).
Now a sequence of secondary actions is a decision (about some future activity A) provided:
\begin{itemize}
\item The sequence of actions causes A to take place (some theory of causality must be assumed).
\item An alternative sequence of actions can be imagined which in a plausible alternative world leads to a future in which activity A will not take place
(the alternative sequence of actions `avoids A'). This alternative must be obtained by modifying actions that incorporate degrees of freedom for
agents involved.
\item No shorter other sequence or subsequence also causes A and at the same time allows for an alternative sequence of secondary actions that avoids A.
\end{itemize}

In complex organizations the formal decision process (thread progressions that qualify as decisions according to the formal criterion) may be so extensive that
it is reasonable to refine their secondary actions recursively into secondary actions  of a first and of a second class. 
Secondary actions of the first class represent the
major formal steps in making a particular decision while the second kind (of secondary actions) may glue together to progressions of actions that at a closer
inspection may be considered decisive (for making the mentioned decision). 

The connection between decisive traces and control codes lies in their sharing of the property that the causalities involved determine their classification
as such. 
For a code to constitute a control code (in the sense of \cite{BMCCL}) its (caused) impact on system behavior is dominant over its self-proclaimed role.
This observation gives room for the suggestion that the theory of control code may be used as a point of departure for development of a similar theory of
decisive progressions that may underly a theory of decision making.
\section{Control code testing}\label{CCT}
It should be noticed that no intellectual approach, that is no form of systematic reasoning,
can exists on which to base the task of understanding or predicting what behavior of a box B is brought about
by it running C, that is by executing B(C).\footnote{The control code need not to have been programmed, 
let alone that it is open source. For the difference between programmed and non-programmed control code see 
\cite{BMCCL} and \cite{Janlert2008}.}  

This observation, in combination with the need to make decisions about control code release, deployment and usage gives
rise to the prominence of testing. This section is aimed at a clarification of some conditions under which control code testing
constitutes a plausible part of control code related decision making activities. The control code is not assumed to be safety critical
in this section. A definition of safety critical control code and an assessment of control code testing in connection with safety critical 
conditions of usage is discussed in Section \ref{SCCC}. In \cite{Jungermann1980} a listing of four `salient variables' concerning 
decision making is given:
\begin{enumerate}
\item possibility of continuity,
\item reversibility,
\item range of effects,
\item time pressure.
\end{enumerate}
These variables are helpful to draw a general picture of the context of decision making envisaged in this section.
It seems reasonable to connect safety critical conditions with reversibility. Thus reversibility of decision is assumed to some degree.
The possibility of continuity is connected with the disadvantage implied by not making use of forthcoming control code. It is assumed 
that this disadvantage is rather high which leads to some but perhaps not acute time pressure. Finally it is important to get an adequate
picture of the full range of effects of the spectrum of decisions that at any moment can be taken about some control code.
\subsection{User side control code testing}\label{USCCT}
The following line of reasoning at the side of a user U may be considered acceptable:
\subsubsection{Acceptance test rule (ATR).}
Let B be a box of type t = t(B). Let C be a CC that is an executable for boxes of type t. 
ATR allows to conclude from the following \ref{NrOfCond} conditions to infer the subsequent \ref{NrOfConcl} conclusions.
\begin{enumerate}
\item C has been manufactured by a trustworthy control code manufacturer M (in particular trusted by U), well 
experienced with boxes of type t(B), and if,
\item the website of M writes that it produces implementations of a service S, which is also written on the 
limited documentation of C, and if,
\item test specialist T is aware of  service S and the utilities a sound implementation may bring with it (for U), as well
as the disutilities that an invalid implementation of S may create (for U), and if,
\item running B(C) on a sample input, as T has been doing,  according to T
indeed provides the service S (that would bring, according to condition \ref{NeedForS} below, 
significant utility UT for U when performed by way of usage instead of test), 
and that compliance is persistent along a number
of runs (tests) with different inputs (which may be provided interactively rather at once), and if,
\item T agrees with the views that M has about Service S and the task of its implementation, and if,
\item S has been automated on control coded devices of type t(B) before, and if,
\item user U of box B trusts test specialist T and if,
\item \label{NeedForS}U needs service S, and if,
\item nobody trusted by U and easy to consult about the matter is knowledgeable about how B(C) technically 
works, and if,
\item  the utility UT of consuming service S to U is substantial, whereas
\item \label{NrOfCond} the disutility of a failed service (intended to provide S) is low.
\end{enumerate}
Then it is plausible for U to perform the following \ref{NrOfConcl} inferences and actions:
\begin{enumerate}
\item to infer that B(C) delivers S on all or most relevant inputs, and,
\item to acquire the right to use C on U's box B, and,
\item \label{NrOfConcl} to decide (conditional of the acquisition of the right of using C for the purpose of providing S)
to make 
use of B(C) when service is is required by (U).
\end{enumerate}

\subsubsection{Rationale for the acceptance test rule.} ATR is an expression of reasoning not included it ATR's 
statement itself. We will highlight this reasoning and thereby explain the presence of each of the mentioned conditions. 
The primary rationale for ATR is that it would be implausibly accidental
for C to deliver S when B(C) is executed. The risk exists that U (or T on behalf of U) is being deceived by M 
but that risk can be ignored because U trusts M. M publicly claims expertise in producing control code for service S
on boxes of type t(B). This confirms ex post plausibility that C is a product of M. T is able to test for delivery of service S
and knows what is at stake when its judgements are wrong. T knows abut `implementations' of S on t(B) type boxes and in
particular (by keeping track of professional literature) T knows about M's strategy for realizing S and T agrees 
with that view. Thus T is not struggling against its disbelief in the general approach taken by M concerning service S. 
U has nobody else to rely on who is more knowledgeable about the technical  details of what is going on when B(C) is
running. Because U needs S it will rely on T's advices in the matter. U's potential gain (utility of using C)
is substantial, while his loss (disutility when, in spite of the preceding facts C, fails to deliver S on box B). Here ATR may
be overly restrictive. In some cases U would go ahead even if the expected disutility of failure of C to deliver S on B is significant.

\subsubsection{Ex ante control code testing for users.}
ATR provides an argument that is only useful for a prospective control code user who is in need of 
an ex ante confirmation that the plan to make use of C is reasonable. Because in general one assumes a 
wide gap of expertise between user and manufacturer it is plausible that a user works and thinks according
to the rule, provided he assumes the ability of a manufacturer to produce sufficiently errorless control code.

The prospective user will base its decisions on functional tests, which are equally informative about processing times.
The tests
performed on behalf of a user might be termed acceptance tests because U may decide not to acquire usage rights
if T is unconvinced.
Strictly speaking 
ATR belongs to single box control code logic because no other boxes are mentioned. It is, however, 
implicit that U assumes that M has successfully performed tests on C with one or more similar 
boxes B$_1$, B$_2$,... of his own.
The plausibility of the acceptance test rule depends on the existence of multiple boxes and assumptions about
executions of comparable control code made on other boxes.

\subsection{Producer side control code testing.}
The intuitive obviousness of testing for a user is suggestive of the option that testing also works for the
manufacturer in ex post mode of operation (after the code has been designed and produced). But this argument
is flawed. The arguments leading a control code producer to perform tests must be quite different in structure.

A well-known, but by no means generally accepted, argument forcefully promoted in \cite{AmmannOffutt2008} and 
pioneered in Beizer's \cite{Beizer1990} test process maturity model,
suggests that control code testing ex post at the manufacturer's side is not functional but must be structural. The purpose
of structural testing being exclusively to improve or validate the control code production process, and not to improve product 
quality in any immediate fashion.

According to \cite{AmmannOffutt2008} functional correctness must have been assured by adequate engineering methods, 
involving requirements engineering,
stepwise refinement, software design, program translation (compilation), and control code generation. Structural 
testing takes the functionality as given and focusses on the rationality of its implementation in an instruction sequence. 
Structural testing is performed with the knowledge of the structure of the code before its compilation into executable form
(relative to type t(B) boxes). Thus assuming that C is a program (it has come about\footnote{Or rather: the most plausible 
conjecture about C's genesis it that it has been compiled from a preexisting instruction sequence} 
via translating an instruction sequence), 
that instruction sequence can be analyzed (verified, simulated, model checked, submitted to code walkthrough, 
peer reviewed etc.) in order to detect and remove design errors (including the so-called programming errors) 
as well as errors of requirements capture. Structural testing belongs to this family of activities.\footnote{It may be even reasonable to claim that:
in the absence of information about C's production process no ex post testing at
manufacturing side produces any substantial degree of confidence.}
\subsubsection{Release test rule (RTR).}
CC manufacturer M
 may use the following rule before deciding to ship CC to the intended user community. 
Shipping precedes the acquisition of licenses by users because users will be allowed to apply ATR before deciding to 
make use of the CC. The rule is called RTR because it is supposed to take place just before a (new) release
of the control code is issued, while many intermediate tests may have been applied during production.

RTR: Suppose the following \ref{NrOfCond2} conditions are fulfilled,
\begin{enumerate}
\item a box type t is given,
\item C has been manufactured by M by means of a production process well-known to M on the basis of requirements (R) that
were available in advance, and if,
\item these requirements provide a clear set of hypotheses (hereafter called SPEC for specification)
which specify what is expected from executing  C (after successful production), and if,
\item C is a compiled polinseq (say P) such that either
\begin{itemize}
\item P has been systematically reviewed (against SPEC) 
according to procedures clearly documented in M's CC production process description, or 
\item P has been  verified (including model checked) by means of documented techniques well-known to M such that M is 
competent it the use of the tools needed for these techniques,
\end{itemize}
and if,
\item M either
\begin{itemize}
\item has performed a substantial number of (successful) tests on C using boxes of type t, 
 similar to the tests which he expects the prospective 
users to perform to the extent that according to the judgement of M user side test primarily deal with the problems that
may arise if their executing boxes differ from the boxes M has been using for testing purposes, or
\item has performed a number of simulations if boxes of type t have not been and cannot be made available to P,
\end{itemize}
and if,
\item there are no further persons around to whom M has easy (but perhaps compensated) access, 
whose advice concerning 
	\begin{itemize}
	\item C and P and the instruction sequencing notation used for writing P, or
	\item the original requirements R and/or the process of its capture, or
	\item software process used for writing C, or
	\item the validation and testing process, that has been applied, or
	\item expected disutilities at user side when C fails to deliver, or 
	\item expected box types at user side,
	\item competing products (CC's for delivering SPEC according to R on boxes of type t) from one or more other and 	
	independent manufacturers,
	\end{itemize}
should be sought because of their intimate knowledge of one or more of these aspects, and if,
\item \label{NrOfCond2} the business case for shipping C is clear to M, and so is the risk analysis involved.
\end{enumerate}
Then it is plausible for M to perform the following \ref{NrOfConcl2} inferences and actions:
\begin{enumerate}
\item that B(C) delivers S on all or most relevant inputs, for boxes of type t at user side, and
\item to ship C to interested users, and
\item to offer them terms for acquisition and use, and
\item \label{NrOfConcl2} to communicate the recommendation to potential users to make use of C.
\end{enumerate}

\section{Definitions for code controlled boxes}
Using the terminology of \cite{Bergstra2010a} an IDBR (informal description by roles) of a code controlled box 
can precede more formal so-called imaginative  definitions. An IDBR is not unique, many alternatives can be 
conceived.\footnote{In \cite{Middelburg2010a} an IDBR for an operating system is provided, while concluding
that more informative definitions of operating systems are unavailable.} An IDBR at the same time contains more informal meaning
and fewer inter concept relations than a so-called conceptual model in the sense of \cite{WandEtAl1995}.

In the previous section of testing almost no assumptions have been made about code controlled boxes, except that use
and test can be distinguished. In order to present more specific arguments about code controlled boxes a more 
refined definition thereof is useful.

Below a proposal is worked out for an IDBR of a code controlled box. In this definition control code execution architecture
is used as an alternative phrase for code controlled box. In \cite{BPEA} a control code execution architecture is termed
a synthetic execution architecture in order to emphasize that the control code constitutes a true part from which the 
execution architecture has been synthesized rather than an abstraction of some part which is only used for its analysis. 

\subsection{Restating the question}
Before turning to the definition of code controlled boxes in more detail it is useful to state once more and in more detail 
which questions are to be addressed by developing an informal control code logic (ICCL), and why that is supposed to be of
any importance. By introducing ICCL a rather specific level of abstraction is created for discussing computers. At this stage
it is unclear which aspects of computing can meaningfully be dealt with at the level of ICCL. Thus at the same time a level 
of abstraction is created and questions about what may be expressed at that level arise.

The following listing of issues intends to clarify in more detail the specific properties of an ICCL level of abstraction and in
addition it clarifies questions that must arise once such a level of abstraction is taken seriously.

\begin{description}
\item{\emph{Fundamental information hiding for computing.}} Understanding a 
computer as a code controlled box achieves a level of abstraction where no assumption is 
made that a mathematical or a logical theory from which to predict machine behavior is available. A user making use of a 
code controlled box needs to trust someone else in order to gain confidence that this is a meaningful action
because he is unable to acquire any substantial confidence in the available technology by looking into its structure and details,
for the simple reason that this information is hidden. This form of information hiding seems to be fundamental for  modern
computing where the large majority of users users cannot conceivably pretend to understand how their boxes arrive at their
observed behavior.
\item{\emph{Technology adverse computer science.}} I am interested in finding out which parts of computer science,
or rather of computing and usage of computer technology, are accessible for those who do not even start to pretend
to know how control codes have been produced and how their execution works. My assumption is that by analyzing
computing technology in terms of the production and usage of boxes and suitable control codes and by explicitly hiding
any information about control code production technology as a substitute for means to predict system behavior a picture
of computing may be found that is realistic in the sense that it is not based on the erroneous assumption that the 
people involved understand what they claim to understand just because of a long standing but fully 
unwarranted tradition of making such claims. 
\item{\emph{ICCL as an ideology.}} I will assume that as a methodology ICCL refers broadly to a state of knowledge and 
a pattern of reasoning employed by one or more individuals working in information technology who do not accept in 
principle and against all odds of the omnipresence of so-called experts that they can base their
activities and carry their responsibilities on any hypothesis about an intellectual understanding of how computers work.
So what activities and tasks are accessible from a perspective of ICCL?
\begin{itemize}
\item Can one perform network and system administration (NSA) on the basis of ICCL?\footnote{This kind of question can be 
compared with the question to define a health care system where only a limited subset
of people involved are supposed to have any (bio) technological understanding of how molecular code based drugs work
inside a human body. Which health care roles are accessible for persons who insist on not at all having such knowledge?} By including
a paper on instruction sequences in \cite{BergstraBurgess2007} I  have committed myself to a negative answer to this 
question. That was premature because no hint of a proof that ICCL is insufficient for NSA has been given 
in \cite{BergstraBethke2007}.

\item Can one (acting at the ICCL level of abstraction) 
be a safe private computer user sufficiently invulnerable to the problems carried around by malicious control codes?
Can anyone understand at all what malicious control code is at the abstraction level of ICCL?

\item Can control code production processes be managed with ICCL based competences only?

\item Can control code deployment processes at the user side be managed and performed on an ICCL basis?

\item Control code usage provides services to its users: can these services be outsourced by managers whose technology
awareness is not beyond ILLC?
\end{itemize}
Answering these questions is far from obvious. Here is a fragmented working hypothesis that results from the 
assumption that
positive answers on these questions may be found: 

\begin{itemize}
\item A large and coherent fraction of IT can be based on ICCL.
\item ICCL based IT at a research level can be called ``behavior oriented computer science''. I suggest that behavior
oriented computer science supports ICCL based IT in the same way as computer science supports IT.\footnote{Behavior oriented 
computer science must not be confused with behavioral computer science which should refer to that part of computer science
which puts the main emphasis on the fact that human users apply and produce computers so that human behavior transpares 
though computing. Behavior oriented computer science is supposed to base its ontology on behaviors rather than on the underlying
structures from which behaviors emerge.}
\item ICCL based IT is a real competence in the sense that full awareness is required of what aspects of IT an individual, 
who is dealing with that particular form of IT, does not know and is not likely to get to know in any meaningful detail 
before having made the major decisions and having performed the main tasks related to that part of his task.
\item Managing trust  lies at the basis of ICCL based IT rather than developing understanding.
\end{itemize}

\item{\emph{What is a computer?}} When moving towards a definition of code controlled boxes it appears that the concept
of a computer cannot be taken for granted. At this stage I have no answer on the following questions:
\begin{itemize}
\item Is the concept of a computer available to ICCL or is ICCL positioned at a level of abstraction where the 
distinction between
computers and other seemingly similar devices cannot be properly 
made.\footnote{One may compare: (i) important aspects of brain 
science may not
be accessible from the more limited perspective of (behavioral) cognitive science, (ii) technical questions about 
combustion 
engines cannot be properly posed in the vocabulary of public transportation, (iii) architects need to know what a 
building is,
but in some cases they may be uninformed about critical aspects (and the corresponding jargon) of civil engineering.}
\item Is `computer science' as a phrase still connected to the concept of a computer or is it simply another (and perhaps 
somewhat outdated) phrase for IT? 
\item If computer science as a description of a scientific activity is based on the 
concept of  computer then what distinguishes that class of devices form the many other products of electrical
and electronic engineering. If such a distinction cannot  be reliably made: is the phrase computer science to be replaced
by another phrase such as information technology.
\item Assuming that a convincing definition of computers can be provided which underlies computer science (if that
is considered a necessity): is some form of universality of the mechanisms involved the critical factor for this 
definition or is mere flexibility in the dependence of machine behavior from control code sufficient. At present I suggest the
following viewpoint as being plausible:
\begin{enumerate}
\item Computer science is mainly concerned with code controlled boxes that are expressively universal 
according some qualitative concepts of computation. In addition:
\begin{enumerate}
\item Computer science needs a clear story about universality concerning computation. The currently prevailing 
story is the theory of Turing machines. 
\item For Turing machines, however, the distinction between 
control code and data code is quite arbitrary. 
\item The viewpoint that a Turing machine is executing a control code
that occupies part of its tape is unconvincing.\footnote{The concept of a Turing machine as a device made for
executing instruction sequences (programs) is also unconvincing. It is unclear why some tape contents ought to be 
considered instructions and other tape contents data.}
\end{enumerate}
\item A notion of universality for boxes is only meaningful for code controlled boxes. A notion of universality can
be developed in terms of ICCL without any dependence on additional computational mechanisms or principles.
\item Without some claim to universality of its boxes no significant demarcation of computer science  can be claimed 
within 
mechanical engineering (mechatronics), electrical engineering, optical engineering, quantum computing technology
and mixtures of these.
\item When forgetting about universality: 
\begin{enumerate}
\item even if code control is a feature of decreasing importance after all, it won't be easily dropped in a 
computer science context, because 
\item non-code controlled boxes are primarily studied outside computer science,
\item the phrase `computer science' is remarkably resilient against the critique that means and ends have been 
confused,\footnote{As if combustion engine technology could stand for automotive technology.}
\end{enumerate}
\item Universality implies that a code controlled box may contain a control code production environment suitable for
generating control codes for itself.
\item Universality therefore implies that its memory can store 
polyadic\footnote{``Poly'' here refers to the plurality of items available for successive usage.
So polythreading is about several threads operating in succession, while multi-threading refers to several threads operating 
concurrently. Multiprogramming as opposed to ``polyprogramming'' is about producing
a plurality of programs that determine threads which are run in a multithreading regime. `Polyadic instruction sequence'
is used in \cite{BergstraMiddelburg2008a}  because `instruction sequence' fails to have the flexibility of referring to a 
plurality of instruction sequences. Polyadic control code refers to  plurality of control codes that may together fulfill the role
of controlling a box.} control codes which result from its own activity and that moreover
execution of a stored polyadic control code is itself achieved as
the effect of an available primitive action which in its turn is a possible effect or side-effect of control code 
execution.\footnote{This latter
requirement goes beyond the specification of an AnARCH (analytic execution architecture) in \cite{BPEA}.} 
\item At this stage matters become confused. Using Kleene's S-n-m theorem from recursion theory as soon as a 
(code controlled) generalized box can compute all computable functions it also computes a universal 
computable function which 
might be considered an interpreter for stored polyadic control codes. But there is a difference between 
interpretation and execution and the capability of a box to execute rather than to interpret a stored polyadic 
control code cannot be inferred from the S-n-m theorem. 
\end{enumerate}
Of course claiming plausibility for the above points of view is only marginally different from formulating all of this as 
additional questions.
\end{itemize}

\item{\emph{A continuum of definitions.}} In \cite{Bergstra2010a} an exposition of methods of definition
has been proposed. The optimal form of a definition is supposed to be a so-called imaginative definition in the form of 
an LSCD (logical solitary concept definition). Initial stages of definitions are found as IDBR's (informal descriptions by role). For 
a concept given by an IDBR usually many different and competing IDBR's can be conceived and many paths towards more
informative LSCD's can be developed.
Some remarks on this matter:
\begin{itemize}
\item When defining the concept of a code controlled box several difficulties emerge. The definitions given below do not
provide completely satisfactory remedies  against these complications. Instead these definitions may be criticized in several
ways and thorough revisions in subsequent work cannot be excluded because of the inherent difficulties of the issues.
\item A code controlled box is a device which is characterized by a significant dependance of its behavior from
its (polyadic) control code. To define a box in all detail requires providing information about this dependence. But
providing complete information is rather unfeasible.\footnote{The
issue may be compared with focusing on a specific human being while ignoring  the molecular biology of his brain. 
One may `define' a human agent role as a role to be 
performed by a human being while not taking into account on purpose any link between the expected behavior of this 
person and the physiological state of his brain. Specifications for his role as well as expectations about
his behavior need to be provided at a higher level of abstraction.}
\item If someone working and thinking at the level of ICCL needs predictions about control coded box behavior an
additional specification for such predictions needs to be provided. Technical information about the control code 
production process is no substitute for this specification.\footnote{An attempt to clarify the distinction between a mere code 
controlled box and a box where behavioral prediction is enabled has been made in \cite{BPEA}. In that paper 
a synthetic execution architecture describes `what it is', while an analytic execution architecture describes how
a machine may work, or rather how it might conceivably work, with so much detail that behavioral predictions can
be derived.}
\end{itemize}

\end{description}

\subsection{An IDBR for code controlled boxes}
A code controlled box B is characterized by the following features and properties:
\begin{enumerate}
\item B has a user interface with elementary controls like switching on and off, and for activating the features mentioned below.
These controls are called operator controls and they are handled by the operator (operator is considered to be
a role of the user if one prefers to think in terms of users).
\item B features a mechanism of code exchange (for a classification of such mechanisms see \ref{CCEM}).
\item B can be loaded with one or more codes (conceptually known as bit sequences).
\item B has a limited number of modes of operation that can be activated from its user interface where the major modes
can be understood as: producing the behavior expected from executing the loaded control codes and data by box B. At this
stage there is no preferred theory of behavior. In the synthetic execution architectures of \cite{BPEA} (a possible
form of box) only a single 
code is loaded and its (only mode of) execution gives rise to a behavior specified as a process in process algebra. In
the machine function based model of \cite{BMCCL} (another conceivable form of box) a sequence of 
codes can be loaded and a mathematical function returning a sequence of codes describes the behavior.
\item If several codes are loaded a distinction between control code and data code can be made or be sought after. For
machine functions that has been worked out in detail in \cite{BMCCL}.
\item If more than one code is classified as control code the control code is called polyadic. This may depend on the 
mode of operation chosen by operator control.
\item B may be viewed as a tool for executing control codes already known from their roles in the context of other boxes.
Alternatively control codes can be viewed as tools for making B work. Both views may co-exist for the same box and codes.
\item Viewing a control code as a bit sequence, a plurality of control codes can be simultaneously in charge of the 
behavior of a box as well. This is called a polyadic control code.\footnote{A number of instruction sequences allowing
jumps between different sequences has been termed a polyadic instruction sequence in \cite{BergstraMiddelburg2008a}. 
The situation with control 
codes is comparable though different. First of all, in contrast with the notion of an instruction sequence, according to some
the notion of a control code may have the flexibility of consisting of more than one bit sequence. Secondly the internal relation
between different parts of a polyadic control code is unclear. Thus unlike the notion of a polyadic instruction sequence the
idea of polyadic control code is in need of additional explanation.}
\item By executing a polyadic control code a box produces a service. Making use of that service is the most obvious
reason for an operator to initiate that execution.
\item Polyadic control codes may but need not have been produced by compiling polyadic instruction sequences that
serve as an intermediate design stage. At the level of abstraction of a code controlled box nothing is known about
the production process of (polyadic) control codes and no theory is given that predicts the behavior during 
execution from the structure, content or form of the (polyadic) control code.\footnote{If it is known how to disassemble 
a polyadic
control code into a polyadic instruction sequence then using the terminology of \cite{BPEA}
an analytic execution architecture can be used to explain the behavior generated by the formal execution of the
polyadic instruction sequence which at the same time models (predicts) the behavior created by box B when executing 
the mentioned polyadic control code.}
\item B is equipped with one or more control code exchange mechanisms as have been specified above. These exchange 
mechanisms can be directly controlled by an operator. A specification how that works is contained as a part of box B.
\end{enumerate}

\subsection{Analytic execution architectures for code controlled boxes}
The IDBR for a code controlled box of the preceding section is unclear about how the transformation from polyadic
control codes to box behavior exactly works. If nothing is known merely a type is described of which boxes are instances.
If it is felt necessary that a code controlled box provides more information then in addition an 
abstract analytic code controlled box can
be given as follows:
\begin{enumerate}
\item A parsing and syntax analysis operator D (disassembler) is introduced which
transforms polyadic control codes C$_p$ for B into structured data, called polyadic structured control code.
\item Volatile data structures specify data that are maintained by a state during the analytic execution of a polyadic 
structured control code. After each termination of an execution the volatile data are reset  to the same initial values.
Permanent data structures maintain information that remains unchanged by termination of an analytic execution.
\item (Analytic) execution of a polyadic structured control code is mathematically defined by means of 
operational transition rules that make use of structured operational semantics concerning the structured control codes.
The adjective analytic for an execution is use because it is at a higher level of abstraction than the reality it is supposed 
to model.
\item It is required that disassembly followed by analytic execution in the context of initialized volatile and permanent
data structures produces the same behavior as an execution of the original polyadic control codes on the given 
code controlled box. This requirement can only be understood is some mathematical definition of the effect of polyadic control
code execution the given box  is known in principle. The intuition for ICCL is that  although such a definition exists in the 
mathematical universe it is considered to be of no
practical use because it is too complex. In software engineering terms the definition can be considered a hidden part
of the specification of the box.
\item It may be maintained that any code controlled box B must be equipped with an abstract\footnote{The
reason to speak of an abstract analytic execution architecture is because there is no implication that disassembly of
C$_p$ will produce a polyadic instruction sequence, a fact that is assumed in \cite{BPEA}.}
 analytic execution architecture AE$_B$ 
as specified in the above fashion. At least the producer of C$_p$ has it at hand. 
\item In general AE$_B$ is supposed to be hidden from the users of C$_p$ and for all end users served by 
those users as well. Thus users know that executing C$_p$ on 
box B produces a behavior which is given (can be predicted) by an analytic execution of D(C$_p$) on AE$_B$ but they 
don't have any practical means of accessing these predictions. Their only way to obtain such information is by actually
executing C$_p$ on B.
\end{enumerate}
\subsection{Intrinsic universality for code controlled boxes}
If a code controlled box is given by a machine function according to the above IDBR, then it is reasonable to
view this alternatively as a so-called synthetic execution architecture (SynARCH) in the sense of \cite{BPEA}. 
The 
adjective synthetic means that it synthesizes parts that model components of an actual or imagined system rather
than abstractions thereof. Opposed to a synthetic architecture \cite{BPEA} proposes the notion of an
analytic execution architecture (AnARCH). Instead of an AnARCH a generalized version of it AnARCH$_g$ has just
been outlined.\footnote{The AnARCH is a special case of an AnARCH$_g$ where the polyadic structured control 
code is required to be a polyadic instruction sequence and the volatile and permanent data structures are 
represented by means of services (see \cite{BergstraMiddelburg2009}).} The components of
an AnARCH are abstractions of what one expects to find as components of a box in practice. The key advantage,
however, of an AnARCH is that it explains (and for that reason predicts)  the results of execution which 
a SynARCH merely lists in the graph of a mathematical function.

In order to make  AnARCH$_g$ produce a model of the computation of C$_p$ on B the control code C$_p$
must be disassembled into structured control code. Now it may be the case that the disassembler 
projection produces structured control code that
has certain parameters constrained by numerical bounds due to the finiteness of the collection
of executables for the given AnARCH$_g$. For instance the number of codes, their length and the number of
occurrences of different features used, may be limited to certain bounds. These bounds may be linked with bounds
concerning the data structures that span the state space of AnARCH$_g$.

In principle one may imagine that the bounds for an AnARCH are considered to be artificial. Then removing these
bounds may be considered a natural generalization step. Such removals may be performed in different ways. It is supposed that  
a natural generalization of AnARCH$_g$ say AnARCH$_G$ is obtained. If there exists an encoding of natural numbers
in inputs and outputs that can be processed and produced by AnARCH$_G$ and if all computable functions on 
natural numbers can be expressed by means of polyadic structured control code for AnARCH$_G$, then AnARCH$_G$ 
is said to be computationally universal. That qualification is extended to AnARCH$_g$.

Now code controlled box B (a SynARCH according to \cite{BPEA})  is said to be intrinsically computationally 
universal if there is a very plausible AnARCH$_g$ together
with some disassembler transformation D for the executables of SynARCH explaining the behavior of SynARCH relative to
some AnARCH$_g$ such that a very plausible general version AnARCH$_G$ found by removing (artificial) 
numerical bounds on the features used in D(C$_p$) has universal computational power.

\subsubsection{Definition of a computer.}
A computer can be defined as a code controlled box which has intrinsically universal computing power in the 
sense just explained.

This definition is ambiguous in the sense that it depends very much on what is considered natural. Further if 
structured control codes have been made up for the purpose of motivating that a given box B is classified as being of 
universal computational power no subsequent notion of naturalness for removing its numerical bounds is easily put
forward in a convincing manner. But the definition can be applied if it is assumed that it is used only to demonstrate that a
given box (intrinsically) has universal computing power (when considering a specific proposal for its generalization by removal of 
one or more finite bounds) 
while it is not used in cases that a positive judgement  that matter is 
refused (that is, no unsuccessful quantification over natural generalizations is assumed, only the lack of such a proposal is reported).

\section{Control code risk analysis}
Both the user and the  producer of control codes can be confronted with risks. Risk analysis may but need not
be invoked as a part of decision making. I will try to draw some picture of decision making and how it may occur in connection
with decisions about release, deployment and usage or control codes.

Needless to say risk analysis is a large subject by itself but an attempt to highlight aspects of primary importance for the
production and use of control codes is legitimate. 

It seems to be the case that user side risk analysis is only possible on the basis of a strategy of use which 
must be known beforehand. If a control code cannot be upgraded after an operational failure thereby `solving the problem'
the context clearly
differs from a context where that can be done. If a user coordinates the activities of many end users the issue of 
risk communication appears which will otherwise trivialize. 

However, providing a full survey of operational models for the use of control codes is unfeasible. After some general remarks 
about risk analysis I will consider the user perspective from the assumption that the user will put the control code at work
to serve a community of end users. Thereafter the producer side perspective will be discussed from the perspective of
a producer which provides control codes for the same type of system. The producer is supposed to have a production
workshop in place which keeps producing improved and modified polyadic control codes for a rather narrowly defined
and coherent class of systems\footnote{Here `system' refers to expected behavior of a box executing the PCC.}

\subsection{Control code related strategic risks}
Both the user and the producer may have control code related assets. For the user this is the control code itself plus 
the result of investments made in preparing its use. For the producer the assets include the production capacity for these 
control codes
as well as knowledge about it, knowledge about competing products, knowledge of actual and potential clients, 
and reputation amongst them.

In \cite{GibsonLouargand2002} strategic risks are summarized based on Simons \cite{Simons1999}. It is concluded in
\cite{GibsonLouargand2002} that
Simons' division in three forms of strategic risk applies to real estate. I put forward that with some minor adaptations this analysis of
strategic risk applies to control codes usage and production just as well.
Simons defines strategic risk as: 
\begin{quote}
an unexpected event or set of conditions that significantly reduces the ability of managers to implement their intended 
business strategy,
\end{quote}
and he distinguishes the following risks (where I have drastically shortened and simplified the descriptions):
\begin{description}
\item{\emph{Operations risk.}} The user may fail to organize control code deployment deployment and efficient use. 
The producer may fail to meet his
objectives due to management mistakes and organizational flaws.
\item{\emph{Asset impairment risk.}} The control code may fail to deliver the required service because of 
changing circumstances. It may fail to run on a new generation of boxes. The production process may become outdated with
new regulations or quality control mechanisms becoming prominent. Or key personnel for production may be leaving.
\item{\emph{Competitive risks.}} Competitors may render the use of a control code or the production of 
modifications and upgrades
of it futile.
\end{description}
Competitive risks are not very special for the case of control code except for the fact that in case
control codes are not protected by patents
or copyrights and these are produced for and sold in a general market, such risks at producer side can appear overnight. This is due
to the minimal cost of transportation and reproduction. User 
side competitive risks are primarily related to the user's core business which is more likely than not outside control 
code production.

Asset impairment on the other 
hand is a non-trivial phenomenon which needs to be taken into account. It is important to notice that qualifying
a control code problem as an instance of asset impairment may be controversial. 
Many external observers have classified Y2K problems
as asset impairment problems to be remedied with reparative actions. But many control codes had on purpose been designed 
so that their functionality was only guaranteed until Y2K. The counterintuitive property of control codes is that 
unavoidable depreciation may take place without any change occurring in the control code itself. Because such
depreciation, if coming from known time limits, is extremely predictable it is not justified to qualify it as a risk as risk must be
connected to uncertainty in the first place.\footnote{It is even reasonable to assume that before a systematic and specific 
analysis has been performed the uncertainty of risks cannot be analyzed in probabilistic terms based on generic and known models either. The more 
random an outcome of some process is, the less reasonable is to speak of it as a risk, unless the probability of that outcome is very small
in which case it might constitute an acceptable risk because it lies below some predetermined threshold. The use of terminology of risks is
rather intricate and as it stands it seems to be the case that some risks are not immune against risk analysis. Once mechanisms become known
and probabilities are clarified some risks become usual system properties and for that reason cease to be analyzed in terms of risks. A terminological 
difficulty is that who speaks of `taking some risk' does not by that speech act alone either introduce a risk or discover an already existing risk
or even prove the existence of a risk (according to risk theory). 
Who `takes a risk' usually has already performed risk analysis and is now taking a chance in a context
with known and different rewards. Problems that have been frequently observed cannot be considered materializations of underlying risks. The primary 
intuition of a materialized risk is a course of events that has not yet been observed and which causes adverse consequences. So one may speak of the 
risk of the USA being unable to settle its governmental debts within the next 25 years because it has not happened before and 
probabilities are hard to assign. Risk analysis regarding this matter may clarify consequences with relative probabilities for (business plans of) other
countries or enterprises or even for individual citizens inside or outside the USA, 
but it will not do away with its status as a risk.}

In the sequel of this section operations risks will be discussed in more detail. Because uncertainties are most obvious
at the user side attention is limited to that side.
Risk, or a plurality of risks may exist without U being aware of that fact.\footnote{Rosa \cite{Rosa2010} suggests that
the  risk which exists independently of any observers may give rise to three epistemological stages for agents or groups of 
agents involved: awareness, analysis and assessment (quantification or qualification of probabilities and impact), 
and management (defining and executing a policy).} It is this observation which Rosa mentions as a decisive argument
for a realist perspective on the concept of risk. 

\subsubsection{Risk materialization and risk identification}
A risk is said to materialize if the sequence of events that constitute its `body' take place. Materialization of a risk
need not imply that the adverse consequences (damage) which justified the qualification of an uncertain conceivable
course of (future)
events as a risk are actually inflicted on those who were at stake. If so the risk has materialized with the feared damage
(or with more/less than the feared damage), if not the risk has materialized without damage. The latter may have been a 
consequence of successful risk management.

Once a specific risk has been identified a measure of
severity of the risk yet has to be attributed. Obtaining such information is a major objective of risk assessment.
Risk assessment is an activity aimed towards obtaining additional information 
to the level that probabilities or probability intervals replace mere uncertainties, and to the point that various outcomes
have been valued in terms of their disutility (damage). Risk assessment may be quite straightforward (at least
conceptually) in which case it follows
a detailed and known sequence of steps. Following \cite{Colyvan2008} not all uncertainties can be understood as probabilities
although many authors insist that uncertainty and probability unavoidably correlate. 
Risk assessment may also be much more scientific in nature, involving new and dedicated research
projects, or it may be social. For instance public outrage created from an adverse outcome can be classified as a risk because it is
real. It can sometimes be predicted from an investigation about risk perception in advance. 

Risks may be so hard to
assess that scientific research is not likely to shed light on them soon. In some cases the risk might be high because of the
sever adverse consequences that can be imagined, while assignment of probabilities or any other form of quantification
of the risk is beyond current scientific capabilities. For such circumstances precautionary policies have been developed. But, it is worth 
noticing that as a decision rule precaution is incoherent according to \cite{Peterson2006}. 

\subsubsection{Risk analysis paradox} 
In order for an agent A to determine whether or not  a significant investment in risk 
assessment and in subsequent risk management is needed (or
justified) before making some decision D it is necessary that A knows that such risks (caused by effectuating D) may exist in principle. 
Here is a bootstrapping phenomenon: some faint awareness
of a risk is needed beforehand to justify a subsequent non-trivial risk assessment.

This `paradox' has its counterpart in risk research. 
Although in theory the perfect application of risk analysis will detect novel risks before they can lead to 
danger\footnote{Following Rosa \cite{Rosa1998,Rosa2010} a danger is a risk with a high probability of leading to damage. 
In the light of previous remarks this definition is manifestly
unsatisfactory because precisely the high probability constitutes an important argument against the use of the term risk. Perhaps the terminology 
can be made consistent by adding the category of pre-risks which combines risk and dangers. This issue is about knowledge. If the pre-risk is known
to materialize with a high probability it is a danger and not a risk (and known not to be a risk), if the same probability is high and if that fact is 
not known the pre-risk is a danger and
it is also a risk, but is classification as a danger cannot be performed. However, if the pre-risk has a high probability of materialization and that is
known it may still be considered a risk if the materialization has never before been observed. Thus: (i) known dangers are not risks 
unless they constitute novel phenomena, and (ii) identification of risks
which are dangers as well is the primary task of risk analysis.} and risk research should
deliver the tool for risk detection,  a survey of risk research 
literature gives rise to the presumption that for risk research just as well as for risk analysis practice the awareness 
of a risk needs to be triggered by adverse human experience before systematic research efforts are made in a 
particular direction.

\subsubsection{Risk management paradox} 
A focus on risk analysis and assessment may hide, and even prevent a 
thorough, professional, and state of the art approach to quality control regarding a control code. It is not justified to view
the consequences of quality problems as risks if it is known how to detect and how to prevent these quality problems. 

If risk analysis is preached in the context of control code both control code errors caused by instruction sequencing 
errors (assuming that the control code has been produced by compiling an instruction sequence) and 
control code errors caused by design errors for the underlying instruction sequence should be considered with
hesitation as potential causes of potential risks. These problems may have been introduced by a negligent 
instruction sequence production process with defective quality control. Subsequently paying lip service to risk management is no 
adequate compensation for that kind of negligence.

Consequently: (in)validity risks (that is risks emerging from adverse consequences of control code production errors) should
only be investigated after it has been established that control code production has been performed at an adequate quality 
level. An estimated high probability of control code invalidity may constitute an incentive to ask for 
higher production quality standards from the control code producer. 
Indeed risk analysis for control code usage in a specific case requires the determination of the quality level 
for control code production. Making sure that this quality level has been achieved is primarily a task for the 
producer. If the suspicion exists that a control code defect may directly or indirectly cause bodily harm of any kind as 
a consequence of executing said control code, that control code is considered to be safety 
critical.\footnote{We will deal with safety critical control code in more detail  in Section \ref{SCCC} below.} 
Users of safety critical control code are entitled at any time to be  informed about
the control code production process for their code. Users need to be able to convince themselves that this production 
process is up to date in principle and that it has been performed adequately in a specific case. 
In the context of safety critical control code user side testing cannot be a substitute for a quality assessment 
of the control code production process, either by means of process inspection or by means of reputation analysis.

\subsection{User side control code risk analysis}
Control code usage can lead to significant and unexpected disutilities which are in some cases best viewed as 
materializations of risk. Risk can be defined 
in many ways. I will now follow Rosa \cite{Rosa1998}, subsequently elaborated in \cite{Rosa2010}, who defines risk as:
\begin{quote}
Risk is a situation or event where something of human value (including humans themselves) has been put at stake and 
where the outcome is uncertain.
\end{quote}
The objective of this definition is to make risk a property of the real world, independent of subjective judgements. Thus 
following Rosa I will assume risk  to have metaphysical content rather than epistemological content. 
Taking a less principled perspective Althaus \cite{Althaus2005} provides a survey
of diciplinary contexts from which the concept of risk may be understood. When U decides to make use of 
control code C for some specified period of time some uncertainties arise:
\begin{enumerate}
\item Will there be any forthcoming execution of control code C on box B initiated by U which has unsatisfactory 
consequences and where
the problem should be diagnosed as a defect of C (rather than a defect of B or rather than as a mistake made by the operator
in deciding to execute C on B at that moment and for that purpose).
\item If there will be a forthcoming execution of C where due to a defect of C `something of human value' 
was at stake and was damaged: is a listing of these matters of human value known in advance (that is when U decides to
acquire the right to use C and decides to start using C) or is there an aspect to this damage that
has in no way been foreseen.
\item If an unforeseen consequence of executing C (considered non-defective until that moment) takes place and this
consequence is considered `severe', will it be the case that retrospectively (ex post) U would have been well-advised to 
put more efforts in risk analysis (ex ante).
\item If the moment arises that U concludes that a more intensive and effective risk analysis for the usage of C for
its own objectives had been necessary: is it reasonable to assume that a stronger effort to detect the risk that eventually
materialized in the form of damage in advance would have been successful.
\end{enumerate}
In the absence of conclusive information about these matters U must in any case ask  the question which damage might be caused by 
various conceivable defects in B and/or in C when execution takes place.\footnote{Here we assume that a defect in B or in
C or in their alignment can (but need not) cause a risk R (or a plurality of risks). R in its turn may or may not materialize 
(thus leading to damage and harm), 
because several other factors not connected to B and C with unknown probabilities
may be involved in the process of the risk R becoming a threat, a danger, and finally, a calamity.} If U sees no potential damage, or only
risks with low severity he may decide not to embark into a subsequent and cumbersome risk analysis.

\subsubsection{Risk levels for control code deployment.}
The following risk levels for the deployment of a control code C can be distinguished:
\begin{itemize}
\item safety critical (risk to health and life of human beings),
\item business critical (risk of enterprise going bankrupt),
\item annual profit critical (risk of writing red figures),
\item employee satisfaction critical (risk of loss of employee satisfaction),
\item customer satisfaction critical (risk of loss of customer satisfaction),
\item customer loyalty critical (risk of loss of customer loyalty),
\item estimated amount of money critical (risk of loss of amount X)
\end{itemize}
To each of these risk levels a corresponding and equally named quality level for control code production is coupled and
a minimum required product quality level can be determined. Therefore user
side control code usage risk assessment can be replaced by the following consecutive steps:
\begin{enumerate}
\item spotting one of these levels for each potential usage (of C on B) that may take place under the
responsibility of U, and determining an upper bound H of these levels for the full range of conceivable usages of C (by U).
\item checking that C has been produced with corresponding quality level corresponding to H or with a higher quality level.
\item performing a theoretical analysis that service S as specified in advance (on the basis of requirements capture) 
to be what C provides when executed on B will not be performed in such a way that damage corresponding to its
risk level can occur. 
\item if all of these steps succeed: infer that risk analysis for C by U has been successful (that is: has not revealed any 
problems). This suffices to conclude that condition \ref{NrOfCond} of ATR of section \ref{USCCT} is met to satisfaction, 
which in turn brings nearer to completion a successful application of ATR.
\end{enumerate}  

\subsubsection{Heuristics for control code usage risk analysis.}
For the specific case of U performing a risk analysis concerning his use of control code C on box B for a range of 
objectives the following assumptions may be taken into account:
\begin{itemize}
\item There is no need to apply the precautionary principle to its full extent. Scientific research will be able to provide an
adequate cause effect relationship concerning the consequences of defects in C. However, that may be overly expensive in
which case the precautionary principle may still be used as a source of inspiration for policy design. 
The specific form of argumentation (and in particular the assumptions on which it is based)
 that makes use of the precautionary principle should be compared with 
Peterson's impossibility proofs in \cite{Peterson2006} in order to guarantee that the argument that one actually makes use of 
is not flawed.
\item Because the risks if any are dependent on the specific context of use, only U can provide the basis of a risk analysis for
his use of C. It is impossible for the manufacturer M of C to perform such a risk analysis once and for all for every potential user
of C. For that reason M cannot ever assume responsibility for the consequences of risks that have unfortunately materialized in
connection with all users of his product C (provided there are many such users and M is unaware of their specific circumstances).
\item Only if C has been manufactured, tailor made, by M for U, with a clear conception of all relevant use cases in advance,
M can be asked to perform a risk analysis to such an extent that U is relieved from that task.
\item On the basis of a sufficiently large collection of conceivable use cases U can decide to what extent a systematic 
risk analysis
is required before deciding to acquire C and to start using it. Perhaps C needs to insure himself against liabilities in 
connection with this usage. Obtaining an insurance policy at a reasonable price will require that U has performed
a substantial risk analysis.
\item Risk analysis is independent from the question to what extent C is correct. Correctness means that its execution on B
will produce a thread or more generally a process which complies with a specification that U has provided or has agreed with.
Non-compliance with the specifications need not constitute a risk, and compliance with these specifications need not
imply the absence of  risks.
\item If U executes C on B to provide services to clients from a client base CB, then agents from this client base must be asked
for assistance in the first bootstrapping phase of risk assessment: each client may in principle provide use cases each featuring
a sequence of events leading to adverse consequences, that is a scenario which is suggestive of a risk.
\item The formidable size of the literature on risk assessment renders it very unlikely that a software engineer E can perform a
professional risk analysis without systematic preparation in the field of risk analysis. However, having performed the risk
 analysis in a 
non-professional fashion constitutes a risk in itself for U. A significant time investment for reading about risk assessment in
general is required for a non risk specialist E to prepare for this task.\footnote{This investment is needed 
if the risk must be managed that (i) E's performance as a risk analyst  is claimed to 
be problematic because of an unacceptable lack of knowledge of the field of risk assessment, and (ii) in case of damage
these questions about the quality of the risk analysis process as conducted by E cause additional liabilities.

There are many books and hundreds of papers on risks and risk analysis but this size of the state of the art is no indication 
that in the case of control code risks the analysis must preferably be put in the hands of  specialized risk analysts who may have had their
practical training in quite remote fields such as nuclear waste disposal or infectious disease control. Two months of full time work
at an academic level of competence should suffice to get a sufficiently clear picture about
the variation of philosophies on the subject of risk theory and on how a number of well-known and 
state of the art risk assessment activities have been performed.
Having made such preparations I guess that indefensible amateurism can be avoided when performing a control code 
risk assessment.}

\item The most prominent ``risk'' that resides entirely within the field of control code production and usage as 
such is the risk that B(C) produces a service which deviates from its specifications. 
But this phenomenon is well-known up to the presence of 
reasonable statistics about it. In the absence of unexpected consequences of an execution failure (implied by ``usage as such''),
this problem is about finding and repairing errors in control code which should be classified under quality management
instead of under risk management.
\item The second most prominent ``risk'' concerns the deviation between specification that were made in advance of 
production of C and requirements that led to those specifications.  Dealing with this problem is called validation and 
again it is not about risk but it is about ordinary quality control.

\end{itemize}
\subsection{Producer side risk analysis}
Providing a complete survey of scenarios with corresponding risk descriptions
for the control code manufacturer is a massive amount
of work (if it can be done at all) and for that reason it is outside the scope of this work. One may however make some 
assumptions about the business model of the control code producer which then provide a point of departure for an initial risk analysis. 
Here are some assumptions:
\begin{enumerate}
\item M accepts no liability from individual users, even if they have paid for the acquisition of a polyadic control code which M 
has released and sold as an implementation of certain tasks.
\item Each polyadic control code C$_p$ shipped by M is the output of a production line which has been set up in order to
produce a steady stream of improvements that take care of new circumstances and modifications which take care of different
user requirements.
\item M is dependent on his reputation in the market. If that reputation is lost no further control codes can be 
sold (or far fewer). Market share for PCC's for the specific family of tasks at hand is of critical importance for M. Below
a certain market share the production line cannot be maintained because new customers will not trust that it will stay in 
business at least as long as their use of a new PCC (including its succession of enhancements and upgrades) has 
been planned to last. Reputation depends on several aspects:
\begin{itemize}
\item User satisfaction for a vast majority of users.
\item Speedy release of upgrades when errors have been detected,
\item  All users profit from the detection of errors in C$_p$ and its variations as discovered by all other users. Users 
obtain regular upgrades that include repairs for problems that have been spotted by the entire user community. 
\item timely release of announced improvements and enhancements.
\item Portability of C$_p$ to a range of different platforms (that is: box types).
\item A buyer U of a PCC C$_p$ will create a vendor lock in for some time and within that period U is dependent on the 
sustained existence of M or at least of the group within M which takes responsibility for C and similar output. M needs to maintain U's 
confidence that the code can be trusted.
\item Sustained marketing potential: M needs a way to communicate to a large audience which PCC's  it has produced and 
how to acquire the rights to make use of those PCC's. This potential is a constituent of reputation. Unless this reputation is available
a new product cannot be sold to users who then know that they will depend on the fact that sufficiently many new 
customers will be found so that the business unit producing C$_p$ like control codes will survive adverse times for M.
\end{itemize}
\item M is aware of other PCC's that different U's may execute on their boxes. P knows that combined or even
sequential use of C$_p$ and other PCC's (for different objectives and from different manufacturers) can
give rise to
so-called security risks to such an extent that only the release of a fast upgrade of C$_p$ can limit the potential damage 
to M's reputation.
\item M is systematically experimenting with such other PCC's from al plausible sources.
\item M is aware that malicious control codes may be imported into user's boxes and that in particular if 
a malicious code enters a box during an execution of C$_p$ (in some cases even essentially facilitated by the use of C$_p$) in such a way
that this phenomenon is repeatable, M will be held accountable for some damages that users are exposed to because of this mechanism.
\end{enumerate}
All of these assumptions give rise to clues for risk analysis. M will need multiple criteria decision making in
order to manage the combined risk that emerge from the risk analysis concerning each of these items just presented. In 
particular security risks are real in the sense that whether or not vulnerabilities exits in principle and to what extent 
these vulnerabilities can and will be exploited and the time scale for such events is quite hard to predict so that an assessment in
terms of probabilities for such fact and events is nearly impossible.

In the light of current computer technology an assessment of security risks requires in depth knowledge of all technology 
involved including the mechanics of control code execution. This is a weakness of the overall organization. Much is to be
gained if security can be analyzed from an ICCL perspective because in that case arguments are simpler and 
mistakes are less likely to be overlooked.

Concerning rule RTR all that can be said is that risk analysis for the producer can be a very complicated effort which finally
leads to a yes or a no, which can be used to satisfy the relevant conditions of RTR.

\section{Control code related quality management}
Given the formidable importance of quality management in today's computer industry the question arises to what extent
quality management can be analyzed and assessed on the basis of an ICCL perspective. This raises
fundamental questions of priority and ordering:
\begin{enumerate}
\item Is the presence of a sound risk analysis for the decision to acquire and to use a polyadic control code 
(that is an application of ATR) a component of process quality, or is conversely, lacking quality a risk that needs to be 
noticed and subsequently assessed. (In other words: is at the user's side quality management in charge of risk management or the 
other way around.) 
\item Is the risk that a producer fails to provide quality code a relevant concept, or is the removal of that risk implicit in the very 
definition of quality code.
\item Is compliance with requirements and specifications an aspect of product quality or should that be presupposed 
before any quality judgements are made. This is a consequential matter: by labeling something as belonging to the
responsibility of quality management its importance is undeniably downgraded. 
\item The alternative, and quite attractive, view is that quality measures aspects which have not been laid down in
requirements and specifications.
\item It is reasonable to define the distance between two PCC's as the smallest number of edit 
operations (inserting or deleting a bit, inserting or deleting an empty control code) which allows to transform one of the PCC's into
the other one (and symmetrically of course). An error (sometimes called a bug) in a control code can be 
understood as a significant deviation of the behavior it generates from intended (specified) behavior which can be remedied
by means of a modification that changes a PCC to a variation of it which is near in terms of this notion of distance.

With some effort sequential composition of such transformations can be described which requires that they don't interfere.
All of this can be made explicit in the theory of string rewriting. Using these concepts it is reasonable to assign 
(in some cases) a number of errors to a PCC, or a number of errors that has been found and repaired (or that number in
a recent period and so on). 

From some stage onwards the statistics of bugs (detection and repair) is considered relevant for product quality 
assessment. Independent from a specific product or product line that statistics is considered relevant for general
production process quality assessment. Following \cite{AmmannOffutt2008} the main virtue of maintaining bug statistics is 
connected with this latter purpose.
\item There seems to be an remarkable agreement in the software engineering literature that product quality for control codes 
cannot be properly assessed without simultaneously looking into the production process thereof. If that is true product quality 
cannot be assessed at an ICCL level of abstraction.
\end{enumerate}

Without a clear position on what it means to produce control code according to requirements and specifications it is difficult
to determine the stage from which the product or one of its versions exists and further efforts are directed towards bug 
elimination and quality progress. 

The question whether quality management or risk management should take the lead cannot be answered in general. It seems
more useful to consider a hierarchy of risks in advance with some `well-known' risks to be dealt with by means of 
quality management and with an ultimate risk management category that will always take priority over quality management
if it is at all applied.

\section{Control code related business cases}
Both ATR and RTR call for the judgement, as encoded in the conditions to the rule, that the agent taking the 
decision has an interest in doing so. To validate that interest a business case must be provided. Writing a survey of 
business cases for control code release and control code deployment in general is not the purpose of this section, it
constitutes a challenge by itself, but some remarks about it are in order.

\subsection{Business case development for control code acquisition and use}
In the rule ATR that describes how a test can lead to a decision the condition is checked that U needs the service S provided 
by executing C on B. This can be validated by  demonstrating  a convincing business case for the use of S.
The business case must cover the cost of 
acquisition (including tests), it must predict the required profits, and it needs to be sufficiently attractive not to be worried
about residual risks that have been detected and subsequently assessed during the risk analysis phase. The business 
case may range from trivial if no additional costs are expected for a well-understood service to quite complex and even 
speculative. It is important that U is aware of the form of his business case. A complex business case needs to incorporate
its own risk analysis. Here we will outline the ingredients for a business case in a complex setting:
\begin{enumerate}
\item U will have a variable number of end users for S as well as a reasonably stable number of support staff.
\item By putting S in place previous methods of working will become obsolete. U needs a description of the initial setting
(stage zero measurement) such that the cost and problems connected with what will become obsolete can be estimated.
\item U needs to: 
\begin{itemize}
\item determine the composition of the required support staff (numbers, competences, organizational structure). U may
need external consultants to help in this phase. At this stage U may need a temporary license for using C that permits
use for fields tests (experiments about the business case) but not for actual usage. In this stage the required 
performance of the support organization needs to be specified (e.g. the maximum number of outstanding calls for the helpdesk,
the average time to solution for an outstanding call, the means of escalation when call handling is taking too long in 
a specific case, the means of escalation employed when overall performance is inadequate).
\item make a plan on how to recruit the support staff, how to set up a management structure and how to provide training.
\item develop a communication plan to the end user community in U's organization and to make a plan on how to get started
and make progress with using S. The plan may involve setting up user groups and expert user groups, end user involvement
in support staff evaluation. Develop a plan concerning the most plausible contingencies (server failure, 
network failure and so on).
\end{itemize}
\item All expected costs of the preceding stages need to be clarified.
\item The advantages of using S in a stable situation need to be compared with the cost and disadvantages of pursuing
old processes. At this stage both quantitative and qualitative data must be combined.
\item A judgement about the business case needs to be stated, taking the expected costs of acquisition and user 
side testing into account.
\end{enumerate}

The above plan is applicable within a single  organization. More complex cases are found when U needs to offer service S
to many independent customers from the public whose interest needs to be created by means of marketing activities. 
Simpler cases within a single organization are obtained by deleting stages and phases from this description.

This section has been included because without a convincing business case no conclusion from test can be drawn 
using ATR. Writing a comprehensive statement about business cases for control code acquisition is not the
objective of this section and it will require much more space. An informative text 
about business cases requires its inclusion in a more comprehensive explanation of the software process. A concise
statement of that nature is contained for instance in Kruchten \cite{Kruchten1996}.

\subsection{Control code autonomy}
A vary remarkable fact about control code in practice is it that it has an autonomous
existence of its own almost independent of executing boxes. 
I will use the phrase control code autonomy to refer to this independence of control codes. Probably control
code autonomy can only prevail if at least some boxes allow exchangeable control codes. 

In order to be sufficiently precise I will assume that boxes are developed in an evolutionary process. Boxes move through generations.
Between generations differences can be significant and may depend on the logical architecture of boxes, whereas within generations differences 
may be high in terms of performance and in terms of other measurable properties but are moderate in terms of architectural structure.

In more specific detail: control code autonomy comprises the following aggregate phenomenon:
\begin{enumerate}
\item   Manufacturing, distribution and ownership of PCC's\footnote{I will often write CC where PCC would perhaps be more appropriate.
As written earlier the notion of a CC may well have the flexibility of consisting of a plurality of codes (bit sequences) in which case being 
explicit about PCC's is overdone and the default (of CC) could be PCC rather than CC with solitary CC as an indication
for a CC consisting of a single bit sequence only.}  is to a large extent independent of 
manufacturing, distribution and ownership of boxes. CC's have a two dimensional classification, one dimension 
representing the functionality (service) a CC is supposed to provide, the other dimension being the box type the 
instances of which it is supposed to control. 
The intuition is that boxes of some type are `universal' for a range of CC types. 
\item More precisely independence offers the possibility of intra-generational inter-box transfers of CC's. 
\item To a much lesser extent CC independence offers inter-generational inter-box transfers of CC's. Inter-generational
CC transfers are  mainly observed 
for CC's that are (and have been produces as) compiled polyadic instruction sequences and it requires that the CC's can be easily 
reengineered into polyadic 
instruction sequences. Inter-generational inter-box transfer represents a case of portability as formalized in \cite{BMCCL}.
\item Rather than relying on inter-generational transfer for keeping their CC's operational, CC's are adapted 
(by their manufacturers) to new box types as soon as these are introduced 
on the market. Adaptation leads to dedicated CC's. The use of dedicated CC's is restricted by license agreements and 
procedures.
\item Just like box types, CC types have a line of development. CC types may be a descendant of one or more 
ancestors. If a CC has no ancestors it is original. Descending chains for (important) CC types may extend in 
time far beyond the customary 
period of usage for an individual box, and may range over the life time of several box generations.
\item At closer inspection even with technologies that offer customers only intra-generational CC exchange the lines of 
development for CC types refer to polyadic instruction sequences (see \cite{BergstraMiddelburg2008a} for 
that notion), from which dedicated control codes are generated by means of (backend's) of (equally dedicated) 
compilers (so-called cross compilers).
\end{enumerate}

Control code autonomy is a remarkable concept. Here are some further observations about it:
\begin{itemize}
\item Ordinary printed
books or newspapers typically don't feature  control code autonomy,
where control code might be an electronic form of the data made accessible by the book.
The development of eBooks, however, which made a significant start on the consumer market
about 2008/9, will lead to control code 
independence for books, admitting for intra-generational  transfers, 
by way of a strict separation between rendering technology and the textual data to be
rendered for reading by a human user.

\item Standardization of CC's for eBooks (e.g. via XML dialects) is necessary to enable vertical CC transfers as well.
\item CC independence made its start as an omnipresent feature of mainstream technology with mainframe business
computing and is now also a dominant feature for personal 
computing. But in many embedded devices (photocamera's, clocks and watches, mobile phones, GPS equipment,
cars, (microwave) ovens and other kitchen equipment, radio's, TV's, other AV equipment) CC independence is not a 
dominant feature (at least at the time of writing this paper). 
\item Every code controlled piece of equipment (box) that needs  to offer only  a fixed number of services, 
known in advance can 
(but need not) be replaced by a (potentially cheaper) piece of technology which is not code controlled. 
\item As it stands code controlled systems design is conceptually prior to the design of non-code controlled systems,
at least beyond a threshold level of complexity of the system.
Non-code controlled systems are simplified products derived from code controlled prototypes that constitute a crucial
stage of development even when code control is intentionally rendered redundant in a final design stage. 
Lines of descendance 
for IT design are primarily visible at the level of prototypes, whereas the line of development for equipment design 
can be followed by visual inspection of final products.\footnote{For microprocessors, however, the situation may be 
comparable with control codes.}
\item It is tempting to compare machine types with biological species and CC autonomy with one of their features, 
for instance the ability to swim. In a succession of generations the ability to swim can appear and disappear and 
reappear depending on the 
ecological context. I probably cannot be asserted that by itself the ability to swim represents definitive progress, 
but it can perhaps 
be maintained that for overall evolution to be imagined retrospectively 
it is essential that in some stage the instances of
some species (which are ancestors of today's non-swimming species) were able to swim.
\item Carrying the previous remark further the development of computing machines may be considered an instance of
evolution with CC autonomy representing both a potential evolutionary advantage and a potential evolutionary
disadvantage for the survival and further development some machine type. This perspective renders the contemplation
of a single machine representing all of computing pointless. The production and perhaps also the use of that machine will call for other 
probably more sophisticated and high powered machines, and the hypothesis is plausible that always some machine types, 
which may or may not occur with low frequencies, must feature CC independence.
\end{itemize}

Control code autonomy is very visible in computing practice, from around 1960 till today (2010) and there is no 
sign that its role is diminshing. The observed occurrence of CC independence in current technology raises a 
number of questions.
\begin{enumerate}
\item Assuming that CC autonomy is a feature provided by a box seller to its customers at non-zero cost: why 
does it survive at all?
\item What explains the important role of CC autonomy for a range of technologies (for instance the laptop), and why is it so much less
important for a number of other seemingly very similar technologies (such as the digital camera)?
\item To what extent is the business case for CC autonomy connected to a market where many boxes of the same 
type will be sold and where many instances of the same CC type are simultaneously in circulation?
\item To what extent is the business case for CC autonomy connected with rapid development in box 
production technology?
\item Is CC autonomy at prototype level (that is during design stages) an essential feature needed for the 
design and development of complex systems (that is even for systems that don't feature CC autonomy)?

\item Which implications for control code production methods derive from an objective to create, to guarantee, or to preserve
control code 
autonomy?\footnote{In \cite{BergstraMiddelburg2008a}  a collection of instruction sequences 
allowing jumps in to other instruction sequences 
within the collection is referred to as a polyadic instruction sequence (polinseq). Polyadic instruction sequencing refers to 
writing or designing polinseq's. This phrase is used instead of the common phrase programming in compliance with 
the view of \cite{BergstraLoots2002} that programs represent polinseq's.

Compilation (with an emphasis on the code generation phase) is the process that automatically translates (projects in the
terminology of \cite{BergstraLoots2002}) a polinseq to a  polyadic CC. Typically compilation is machine type dependent and it
is performed by a code controlled machine itself.

I propose the \emph{polyadic instruction sequencing hypothesis} which states that a technology that features CC autonomy 
makes use of CC's which result from
compilation of polinseq's. Further the design and development of codes takes place in terms of polinseq's rather 
than in terms of (executable) polyadic CC's. 

The \emph{high level polinseq hypothesis} adds to this that an evolution of CC's making use of the design of polinseq's will 
stagnate if only the bare notational format of \cite{BergstraLoots2002,BergstraMiddelburg2009} may be used. Instead it 
is to be assumed that
a systematic evolution of polinseq's can only take place if these are themselves found by way of compilation from 
higher level (that is more abstract or generic) notations (so-called high level program notations).

Taking this further it may hypothesized that different high level program notations are engaged in a competition 
between their owner/designers. This fierce competition is needed to make sure that the CC market place is so active 
that the additional costs of CC autonomy are compensated (for sufficiently many users) by its advantages.}

\item Is the emergence of CC autonomy an evolutionary phenomenon, in the sense that a technology 
making use of
CC autonomy is more competitive than a technology without CC autonomy?  Or can we merely say 
that, given the survival of technology lines that offer CC autonomy in each generation, CC autonomy 
is a sufficiently strong feature
not to trigger fast elimination by other technologies that don't have it on offer. 
\item If the latter is the case, is such a situation
still in need of an explanation, or is it simply sufficient to point to today's IT marketplace and to assess that CC 
autonomy is still a common feature. In the latter case the question whether or not a killer technology for CC 
autonomy 
exits or can be developed is ignored and one is satisfied with the observation that such technologies have not yet been 
developed to a sufficient level of proliferation to threaten the survival of CC autonomy featuring product lines, without
making any prediction about the future of CC autonomy as a technological strategy.
\item Even if CC autonomy is featured by equipment used for machine builders and designers that does not imply
that CC autonomy is in any way visible for a larger user community. Is it conceivable that CC autonomy 
will become a niche phenomenon unknown to large user communities. This would apply already if all downloads of new 
CC's are performed automatically from the internet but without introducing new functionalities at least from a 
user perspective.
\end{enumerate}

\subsection{CC autonomy and the single box case}
Assuming the presence of only a single box already introduces several scenarios that can serve as a business 
case for CC autonomy from a user's perspective. Here are three examples of such scenarios.
\begin{itemize}
\item \emph{CC development and subsequent execution on the same machine.} User U may imagine a new 
functionality. When a polinsequensing environment is available on his machine he can develop the application 
himself on his machine. The CC
is available as data. As a CC he can execute it on his machine, as data he can exchange it via the internet. In this setting
it would be an unreasonably artificial constraint to maintain that U's machine is not code controlled. If so that is probably
due to strict security mechanisms which actively block CC transportability into U's machine. In any case U works 
as if the CC's are independent.
\item \emph{Downloadable CC's for remote machines.} Imagine a bored machine user U on the moon unable to produce his
own computer games. 
By remotely downloading a CC for a new computer game he can keep up to date with his children's progress on gaming. 
Possibilities for physical transportation of a CC are minimal, options for acquiring a new non-code controlled game 
console allowing to play the latest game are practically absent.
\item \emph{Downloadable CC's for new functionalities.} Imagine a user U who bought a laptop with internet connection. For U
the concept of internet telephone might have been novel or unknown when buying his machine. But downloading 
Skype allows him to make unexpectedly cheap international connections with cellular phones throughout the world. 
So he is happy that this can be done by downloading a new CC (in this case from Skype).
\end{itemize}

\section{CC autonomy considered a feature}
The simplest conception of CC autonomy for a box type is that it constitutes a feature that the clients of its
manufacturer or retailer have a substantial interest in. Like an open roof for a car, or a built in  CD-player. 
It has been added because customers
are willing to pay for it. In the case of CC autonomy one may wonder why users want to pay for it but that 
may be simply a matter of taste.
Now continuing with the features of cars as an example, hybrid propulsion which may also be considered a feature of a single
car is a comparable feature. Hybrid cars can be used in very 
much the same way as conventional cars with combustion engines only. 

But electric cars (without combustion engine)
are a wholly different matter. For a single 
electric car to exist there is a need for a full scale distribution system for electricity with dedicated charging stations. 
Thus electric propulsion is a feature that cannot be understood in isolation for a single car. 
Similarly lead free petrol consumption is a feature of car engines which may become dominant without any users being 
particularly enthusiastic about it, but its adoption requires that many car drivers participate in its use.
We can formulate some assumptions about CC autonomy as a feature.
\begin{enumerate}
\item If boxes of type t are sold to a significant market and type t prescribes CC autonomy (that is boxes of 
type t are code controlled) than either this state of affairs is known to all users and these have obtained instructions on how to 
make use of it illustrated by
convincing use cases or else this state of affairs is undisclosed to all users, and the option to exchange a major CC in a box B of type t 
may at best result from hacking B, or from covert operations from the side of the control code provider.
\item Most if not all boxes capable of delivering  entertainment services provide CC autonomy.
\item Exploiting CC autonomy from the side of the user can have one or more of three major and different objectives: 
	\begin{itemize}
	\item enabling a box to establish new functionalities (perhaps
	by executing newly obtained CC's that were designed and manufactured after production of the box, and more
	importantly after the user's acquisition of the box),
	\item improving the quality of a service (functionality) already present in B at the time of delivery to its customer 
	(typical example an OS security update for a laptop), and
	\item allowing a user to design, develop and execute a new functionality himself. This requires that box B is equipped
	with a complete programming environment. The latter feature was prominent for classical workstations (popular between 	
	say 1980 and 2000) of it is still present on today's PC's and laptops. But it has been omitted with cell phones and it
	has not been explicitly included into a number of smart 
	phones that were developed on the basis of cell phones although these machines
	have functionalities quite comparable to a laptop.
	\end{itemize}
\item The fact that a computer provides a full fledged instruction sequence production environment capable of manufacturing control 
code for itself may soon be a romantic memory from the past. It is difficult to see why that arrangement constitutes and
advantage on the long run. As instruction sequencing becomes more involved, increasingly complex systems are needed for its 
support. For hardware design it is equally unconvincing that the design or the redesign of a hardware type 
can be performed with the exclusive help of a box of that type. At least for boxes that serve as components of
safety critical systems it is plausible that the
systems used for their 
CC design and development are larger and faster than these boxes themselves. On the long run
 that will hold for most box types.
\item If type t specifies more or less fixed functionalities and the design of type t moves through a quick evolution mainly
focused on improving the attractiveness of the functionalities offered (typical example: digital cameras), then 
improvements in control code are dealt with in the same way as hardware improvements (box improvements). If CC
improvements constitute sufficient grounds for selling new boxes (with fixed CC's) then manufacturers obtain an advantage
from not offering CC autonomy and from selling new boxes more frequently.
\item Assuming that CC distribution can be both quick and cheap a new CC (for instance for a game) can be developed
into a hype which generates substantial profits. Indeed if there is a large basis of boxes of type t and a new 
service S for type t is developed as a CC (say C$_p$) then S can be distributed by merely distributing instances of C$_p$.
This is a very important mechanism comparable to  the development of top-hits in pop-music.

Thus competitive development of CC's is made much more challenging and potentially rewarding when a large base of
CC independent boxes is around. The gaming industry makes ample use of this mechanism. It is reasonable to believe 
that this will not change in the near future.
\item A further major incentive for CC autonomy is CC quality improvement during box life-time. Major examples are OS
replacement and OS updates, including the very frequent security updates. All complex CC's may require regular updates
in principle.  This mechanism forces users into system administration and this unfortunate necessity is a
disadvantage which simply reflects the unstable and poor quality of many CC's distributed today. In office networks 
centralized administration will often take these tasks out of the hands of PC (or laptop) users and by forbidding users to
install CC's at their own initiative (not giving them system administration or superuser rights) the boxes are made
non CC independent from a user's perspective. 
\end{enumerate}

\section{Control code testing for safety critical embedded systems}\label{SCCC}
After a survey of software testing literature  Middelburg  concludes in \cite{Middelburg2010b} that no formalization of software testing as an
experimental process can be found in the computing literature and that no results are known about what information concerning
computer software can be established better by means of testing than by other methods of investigation if such are available.
I assume that these conclusions are valid for control code testing as well, because that can be considered a sub-theme of 
software testing. In this section I will argue that there is a comprehensible rationale for control code testing in the context of 
safety critical embedded systems. This rationale is not dependent on a hypothesized but unproven superiority of testing over other means
of analysis according to whatever criterion.

I will now assume that a prospective user U intends to operate a robotic system R containing at least one 
code controlled machine M and that he contemplates the usage of polyadic control code C$_p$ for a safety critical operation of R.
In a practical context R will probably contain more computational devices each of which are likely to execute several 
control codes but in order to simplify the discussion these pluralities are ignored.

I assume that once started U will be unable to replace C$_p$ if he concludes that it causes problems. Further I assume that
failure of the operation of R  is likely to cause bodily harm to some persons different from U. Thus R is a safety critical system. M inside
R is an embedded computer. It is assumed that R's behavior consists of communicative signals and messages exchanged with its 
environment as incorporated in R. A plausible view is that M is in control of parts of the behavior of R and C$_p$ is in control of the 
behavior of M. How activities of M are triggered and terminated is left unanalyzed.

\subsection{Safety critical system components in detail}
Suppose that U is completely unwilling to perform any theoretical analysis in advance in order to predict what happens 
when R is put into action. How can U arrive at the conclusion that R is safety critical? Only by actually putting R into action and then
observing the bodily harm done, perhaps repeating this until he becomes convinced that the irreversible damage inflicted is 
probably caused by his putting R into action under certain conditions. 
This way of information gathering should not be included under testing. It constitutes a rather drastic form of learning by doing, by modifying
designs containing mistakes after they have surfaced so until their causes can be identified and valid designs can be put in place. 

If U insists to infer mission criticality of R without taking the risk that irreversible damage occurs he will need some predictive capacity
which draws conclusions in the absence of actual operations. This implies that a minimum of theory is needed. Consider jumping out of an 
airplane equipped with a parachute. How to conclude that the parachute is a safety critical system component? This can be done in three steps:
(i) conceptually replace the parachute by a device that will instantly dissolve in the air, (ii) then use the laws of gravity plus known facts about
air resistance to compute an expected speed of ground impact given the totally dysfunctional parachute, (iii) use knowledge about the 
tolerances of the human body to infer that severe consequences are likely.

Thus a component of a system is safety critical within that system if it can be hypothetically replaced by another component in such a 
way that known theory makes the prediction of harmful consequences from a hypothetical operation very plausible. Returning to user U with
his robotic system R containing embedded code controlled device M with polyadic control code C$_p$: U can conclude that M is safety critical within
R if it can be hypothetically replaced by a similar device M$^{\prime}$ with different behavior which is very likely to cause severe damage when put into
operation inside R. These conclusions must be based on some theory about the behavior of R depending on the functionality of M and on some
know relation between problematic behavior of R and expected inflicted damage.

For this definition to make real sense U needs to know in addition that some M can be imagined which, when installed inside R will allow its 
adequate operation thus avoiding harmful consequences. If no such M is conceivable there is no basis for the assumption that M is a safety critical
component of R because the design of R itself is essentially flawed. Obtaining this information is far from trivial. It requires a sound model of R and 
how its behavior depends on various variations of M from which predictions can be inferred. Nevertheless if M is to be considered a safety critical
component of R this second piece of information must have been somehow established, potentially a highly demanding task requiring 
formal models simulations and so on.

This pattern can be repeated when considering C$_p$. This PCC is safety critical for component M inside R provided:

\begin{enumerate}
\item M is safety critical inside R,
\item For some conceivable C$^{\prime}_p$ the prediction that operating R[M[C$^{\prime}_p$]] will not cause severe damage can be made with sufficient confidence,
\item For some conceivable C$^{\prime\prime}_p$ the prediction that operating R[M[C$^{\prime\prime}_p$]] will have 
severe adverse consequences can be made with a 
very high level of confidence.
\end{enumerate}

Each of these conditions can only be established by means of a theoretical account of the systems involved which allows making predictions. Of
course that account may in part need to make use of experimental (that is measured) data, comparable with the parachute case above. 
Only if gravity, as measured,  exceeds a certain threshold the feared negative consequences can be expected and so on.

\subsection{Operating safety critical systems with safety critical components}
It is assumed that at some moment U will indeed decide to start the operation of R. The situation for U is as follows:
\begin{itemize}
\item U must hope for the best.
\item U has a certain amount of confidence that the mission will proceed without serious problems. This confidence is the primary justification of the
decision to put R into action.
\item U's confidence is primarily based on the confidence that none of the safety critical components of R (recursively) will cause failure of the operation of R.
This decomposed confidence can have different grounds for different parts of R: 
\begin{itemize}
\item in some cases it is based on trust concerning the producer of a part 
manufactured elsewhere, 
\item in other cases it is based on a full and formal validation of the design of system components in relation to the context in which 
these will operate,
\item yet in other cases new experimental data concerning system components will produce information needed to complete overall validation.
\end{itemize}

\end{itemize}

Anyhow the prediction that R will work without inflicting damage cannot be produced by mere logical reasoning. Some knowledge about the 
system and its components must have been collected by other means. Altogether the ultimate category that guides U's decisions is confidence
rather than logical proof.

\subsection{Plausibility of control code testing for safety critical control codes}
Given the definition of safety criticality for C$_p$ as outlined above: when will U consider testing of C$_p$ a an activity which increases his confidence that
operating R will not crash on problems caused by C$_p$. This is a matter of hypothetical psychology rather than logic. Nevertheless the following 
sequence of observations leads to a confirmation that control code testing can be a rational way of proceeding for which no obvious alternative 
can be put forward:
\begin{itemize}
\item If U knows that C$_p$ has been produced via instruction sequencing he will prefer to have an analytical execution architecture of M[C$_p$] on
which to base the prediction that this component will not cause any problems within R. This kind of analysis can serve a role as well for
validating the second condition of safety criticality before arriving to the conclusion that C$_p$ is indeed safety critical (for R as deployed in M).
\item Tests consisting of operations of R[M[C$_p$]] in safe circumstances (that is circumstances where failures of R do not have the adverse consequences
feared in actual use) where no problems are reported which can be traced back to C$_p$ will increase confidence of U in C$_p$.
\item A test with a modified PCC, C$^{\prime}_p$ of a system variation R[M[C$^{\prime}_p$]] that develops a failure not present in a test of R[M[C$_p$]]
produces important belief that C$_p$ is safety critical, thus creating justification for a thorough validation of the second criterion that some conceivable PCC
is capable of controlling M in a desired way, and additionally increasing the justification for extensive testing of R with circumstances that require adequate 
operation of M.
\item From this observation it may be inferred that control code testing should include testing of systems that execute control code inside embedded devices.
Nevertheless it seems reasonable to distinguish pure control code testing where the control code is considered a part of a computing device only from
embedded control code testing where the embedding of the code controlled device is also taken into consideration.
\item The more known the behavior that M is supposed to produce is, the more plausible it is for U to derive confidence in C$_p$ from trust in its
manufacturer. 
\item When U is in need to make a best possible attempt for operating R under time pressure because a delay of operating R may lead to damage 
comparable to the damage expected from its faulty operation, it can be reasonable to base confidence in the safety critical control code C$_p$ on a series of 
successful tests in combination with an informal analysis of an analytic execution architecture of M[C$_p$].
\item Tests become more informative as precise specifications of the behavior expected from M and precise information concerning the interface between
M and R is lacking. It such circumstances tests reduce uncertainties about the context of C$_p$ rather than uncertainties about C$_p$ itself. These uncertainties
cannot be reduced by merely considering M[C$_p$] and validating (by means of testing or by means of formal analysis or a combination of these) 
that under control of C$_p$ the device M[C$_p$] works according to some given specification.
\item In the latter circumstances control code testing cannot be replaced by a formalized alternative means of analysis even if the polyadic control code is known
to be a compiled polyadic instruction sequence and an analytic execution architecture for M is known that enables predicting its behavior for a given C$_p$.
\item It seems to rest on some form of general engineering intuition to propose and believe that some C$^{\prime}_p$ can be found in principle such that 
R[M[C$^{\prime}_p$]] will not fail. This is very much connected with a notion of universality of M. Once M is accepted as universal (a matter of trust) and
taking a fairly philosophical perspective on the architecture of R it follows somehow `logically' that an adequate control code C$^{\prime}_p$ exists in principle;
it just needs to be found. 

Remarkably the combination of this philosophical perspective with the assumption of M's universality allows to infer the positive information needed to assert that
C$_p$ is safety critical without much formal modeling work. This is consistent with an informal view of the interaction between M and R and it creates a
context where pure control code testing (which might in principle be replaced by alternative methods of analysis for control codes explained by 
an analytic execution architecture) must be augmented with embedded control code testing (which cannot be replaced by generic formal methods from
computer system semantics).
\item So I conclude that an intuition of universality of control coded devices plus a general perspective on what may be expected from an embedded
digital device creates a setting where control code testing and its embedded extension can play a vital role in the design of safety critical systems, a role that 
cannot easily be replaced by formal analysis or verification, in particular not in the presence of temporal pressure.
\item This conclusion stands even if one agrees that the question raised by Middelburg in \cite{Middelburg2010b} about when pure software testing 
(here identified with control code testing given that control codes have been produced by means of instruction sequencing) is more effective or efficient
than other means of analysis (which are guaranteed to exist on the basis of an analytic execution architecture) not involving the physical process of 
control code execution, is still open (and for that reason might have a negative answer).
\end{itemize}

\bibliographystyle{plain}

\end{document}